# Fixing non-positive energies in higher-order homogenization

MANON THBAUT
Laboratoire de Mécanique des Solides
CNRS, Institut Polytechnique de Paris
91120 Palaiseau, France

BASILE AUDOLY
Laboratoire de Mécanique des Solides
CNRS, Institut Polytechnique de Paris
91120 Palaiseau, France

CLAIRE LESTRINGANT
Institut Jean Le Rond d'Alembert,
Sorbonne Université, CNRS,
75005 Paris, France

**Abstract**

Energy functionals produced by second-order homogenization of periodic elastic structures commonly feature negative gradient moduli. We show that this undesirable property is caused by the truncation of the energy expansion in powers of the small scale separation parameter. By revisiting Cholesky's LDLT decomposition, we propose an alternative truncation method that restores positivity while preserving the order of accuracy. We illustrate this method on a variety of periodic structures, both continuous and discrete, and derive compact analytical expressions of the homogenized energy that are positive and accurate to second order. The method can also cure the energy functionals produced by second-order *dimension reduction*, which suffer similar non-positivity issues. It naturally extends beyond second order.

**Keywords** Linear elasticity, Variational asymptotic homogenization, Higher-order homogenization

## 1 Introduction

Asymptotic homogenization is a rigorous tool to derive the effective properties of a given periodic elastic microstructure [AS77, BP89, BLP11, San80]. It delivers the microscopic strain as a series expansion involving the successive gradients of a macroscopic displacement. The terms in this expansion depend on localization tensors that are obtained as solutions of a cascade of cell problems. At leading order, a standard effective Cauchy continuum is obtained. Pushed to higher-order, the method yields an equivalent continuum whose energy depends on the successive gradients of the macroscopic strain [Bou96, BS11, GK89, LM18]. This correction allows to capture *size-effects* that typically occur when microscopic fields vary on the scale of a few cells [DLSS22, YAL24].

There remains a stumbling block to the use of these higher-order models: in most cases, the effective second-order elastic stiffness delivered by the homogenization procedure is non-positive [ABV16, AL23, DLSS22, LM18, SC00]. The corresponding second-order effective energy is therefore non-convex, which makes the equilibrium boundary-value problem mathematically ill-posed. Typically, spurious solutions having short-scale oscillations are produced by the homogenization procedure [TAL24a]. Besides, the negative stiffness associated with short-scale perturbations leads to severe issues when the homogenized energy functional is minimized numerically [KA24].

Some remedies have been proposed. The so-called 'Boussinesq trick' is often applied to the strong form of the equilibrium, effectively turning the gradient stiffness into a positive one [ABV16]. The trick however introduces higher-order derivatives of the applied load, degrading its smoothness, and introduces quantities whose physical meaning is unclear. Another option is to trim to zero the negative eigenvalues of the second-order stiffness tensor [DLSS22]. This option is interesting in the particular case of structures whose leading-order elasticity is degenerate but comes at the price of degrading the order of accuracy of the homogenized model. Another approach has been proposed recently, on which the present work builds up. It has been observed in [KA24] that the positivity of the original microscopic problem can be preserved by carrying out homogenization at the energy level (and not at the level of the equilibrium equations) and by truncating the microscopic strain appearing in this energy rather than the energy itself. In this approach, positivity is restored at the price of calculating a number of *non-asymptotic* higher-order energy terms whose expressions are cumbersome.

In the present work, we improve on the latter method and propose an original truncation procedure applicable to the asymptotic expansion of the homogenized energy. It is applicable to higher-order homogenization at any order, including second-order homogenization, and preserves the positivity of the original 'microscopic' energy that gets homogenized. In the context of second-order homogenization, the effective energy obtained in this way is the sum of two terms: (*i*) a first term in which the leading-order homogenized elastic moduli are combined with a *generalized strain*, which we derive in the form of the usual macroscopic strain corrected by strain gradients, and (*ii*) a second term in which higher-order elastic moduli are combined with the strain-gradient. The procedure warrants that both the classical (leading-order) elastic moduli and the higher-order ones are non-negative. The proposed truncation method preserves asymptotic correctness: the difference between the positive energy functional delivered by the new method and the non-positive one produced by the standard approach is not larger than the homogenization error. The early version of the truncation method which we proposed recently in [KA24] shares the same key properties but requires calculating a number of 'negligible' homogenized elastic moduli. The truncation method proposed here is significantly easier to implement and yields much more compact energy functionals.

The proposed truncation method is inspired from Cholesky decomposition, a popular numerical method used for solving linear systems of equations involving a positive matrix $\boldsymbol{t}$, see [Str07]. The Cholesky method produces a diagonal matrix $\boldsymbol{D} \geqslant 0$ having positive entries and a lower triangular matrix $\boldsymbol{L}$ having 1's on the diagonal, such that the original matrix can be decomposed as $\boldsymbol{t} = \boldsymbol{L} \cdot \boldsymbol{D} \cdot \boldsymbol{L}^T$. Whereas the Cholesky method is primarily used as a numerical method, we revisit it here in the context of homogenization (see Section 5). Specifically, we apply the first few iterations of the Cholesky method to a block-matrix $\boldsymbol{t}$ representing the asymptotic expansion of the energy, and observe that this produces a truncated energy possessing the desired properties of positivity and asymptotic correctness. The positive, block-diagonal entries in the matrix $\boldsymbol{D} \geqslant 0$ are interpreted as homogenized moduli and the entries in the lower triangular matrix $\boldsymbol{L}$ as coefficients defining the generalized strain.

The approach is illustrated with a variety of periodic structures, including continuous microstructures (Section 2) and discrete truss and beam lattices (Section 3 and 2.4). We also cover the case of non-centro-symmetric structures, whose homogenized energy couples strain and strain gradient (Section 6). In passing, we explain the empirical observation that strain-gradient moduli are never negative in *soft* structures, *i.e.*, when the leading-order elasticity is degenerate [DLSS22, S18a, Dur22]. Finally, we show that our method naturally extends to higher-order *dimension reduction* and fixes the negative higher-order effective moduli that are found in, *e.g.*, the bending of elastic beams (Section 7).

In all our examples, we follow the *energy-based* higher-order homogenization method developed in [Le15, AL23]. The macroscopic displacement is prescribed and the small-scale, oscillatory component of the displacement (microscopic shift) is found by optimizing the elastic energy. The potential of the external loading is ignored in this process as the external loading is coupled to the macroscopic displacement but not to the microscopic shift under standard scaling assumptions on the load magnitude—the external loading can be restored later, when the homogenized model is used to solve a structural problem, see the discussion in Section 3.1 in [AL23]. Next, the solution for the microscopic shift is inserted into the elastic energy, delivering a homogenized energy functional $\Phi^\star$. The present work is concerned with the final step, which is to truncate $\Phi^\star$ in a way that preserves positivity.

# 2 A laminate subjected to antiplane deformation

In this section we analyze a laminate, *i.e.*, an elastic medium made up of layers having dissimilar elastic properties. In the interest of simplicity, we limit attention to antiplane deformation: the layers are periodic in the direction denoted as $\boldsymbol{i}_1$ in Figure 2.1, and the displacement takes place along an orthogonal direction $\boldsymbol{i}_3$ and is invariant along $\boldsymbol{i}_3$. There is no major difficulty in lifting the assumption of antiplane elasticity but the calculations are more involved.

The beginning of our analysis closely follows the work of [Le15], where an asymptotic homogenization procedure has been used to derive the equivalent elastic properties up to second order in the scale separation parameter $\eta$. In Section 2.4, we recall the deficiencies of the homogenized energy functionals obtained in this prior work. In the following subsections, we discuss alternate truncation procedures, including one in Section 2.6 that preserves positivity and is simple to set up.

## 2.1 Geometry and elastic energy

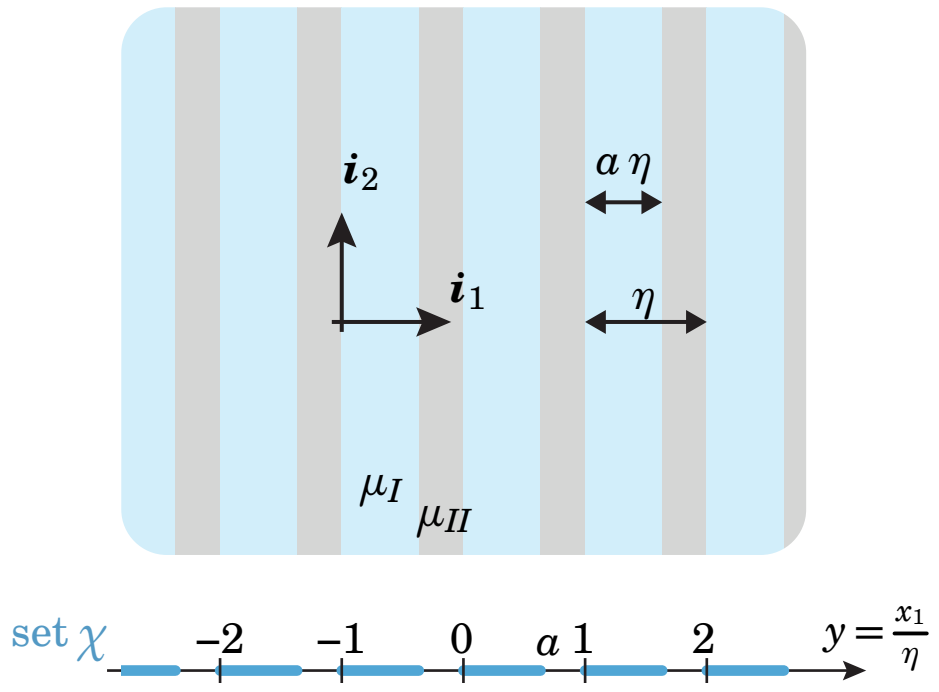

**Figure 2.1.** Geometry of a laminate having periodicity $\eta$ in the direction $\boldsymbol{i}_1$. The laminate is invariant in the direction $\boldsymbol{i}_3$ perpendicular to the plane $(\boldsymbol{i}_1, \boldsymbol{i}_2)$ shown. The shear modulus of the phases (shown in blue and grey) are $\mu_I$ and $\mu_{II}$, respectively. We limit attention to the case of linear, antiplane elasticity: the displacement is along the transverse direction $\boldsymbol{i}_3$, and invariant in this direction.

We consider the infinite laminate sketched in Figure 2.1. It is periodic along $\boldsymbol{i}_1$ with spatial period $\eta > 0$. The phases are made up of a linear isotropic material, with respective shear moduli $\mu_I > 0$ and $\mu_{II} > 0$. We denote as $a$ the fraction of the first phase, with $0 < a < 1$. The domain of the first phase is the set of points $\boldsymbol{x} = (x_1, x_2) \in \mathbb{R}^2$ such that

$$\frac{x_1}{\eta} \in \chi \qquad \text{(phase } I\text{)} \tag{2.1}$$

where $\chi$ is the set of scaled coordinates $y = \eta^{-1} x_1$ covering phase $I$, as shown at the bottom of Figure 2.1,

$$\chi = \bigcup_{i \in \mathbb{Z}} [i, i+a]. \tag{2.2}$$

The set $\chi$ contains the real numbers $y$ of the form $i+b$ where $i$ is a signed integer and $b$ is the fractional part of $y$, with $0 \leqslant b \leqslant a < 1$.

We limit attention to out-of-plane loading, *i.e.*, assume that the displacement $\boldsymbol{v}$ is along the transverse direction $\boldsymbol{i}_3$, and invariant in the transverse direction,

$$\boldsymbol{v}(x_1, x_2) = v(x_1, x_2)\, \boldsymbol{i}_3. \tag{2.3}$$

The in-plane gradient of the transverse displacement $v(x_1, x_2)$ is denoted as

$$\boldsymbol{E}(x_1, x_2) = \nabla v(x_1, x_2) = \begin{pmatrix} \frac{\partial v}{\partial x_1}(x_1, x_2) \\ \frac{\partial v}{\partial x_2}(x_1, x_2) \end{pmatrix} \in \mathbb{R}^2. \tag{2.4}$$

The classical, linearized strain $\tilde{\boldsymbol{E}}$ from 3D continuum mechanics can be expressed in terms of the reduced strain vector $\boldsymbol{E}$ as

$$\tilde{\boldsymbol{E}}(x_1, x_2) = \frac{1}{2}\left(\boldsymbol{E}(x_1, x_2) \otimes \boldsymbol{i}_3 + \boldsymbol{i}_3 \otimes \boldsymbol{E}(x_1, x_2)\right), \tag{2.5}$$

where $\otimes$ denotes the outer product. With $\operatorname{tr} \tilde{\boldsymbol{E}} = 0$, the strain energy $\Phi$ of the laminate, per unit thickness in the transverse direction $\boldsymbol{i}_3$, can be expressed as

$$\begin{aligned} \Phi[\boldsymbol{E}] &= \frac{1}{2} \int_{\mathbb{R}^2} 2\mu(x_1)\, \tilde{\boldsymbol{E}} : \tilde{\boldsymbol{E}}\, \mathrm{d}\boldsymbol{x} \\ &= \frac{1}{2} \int_{\mathbb{R}^2} \mu(x_1)\, \|\boldsymbol{E}(x_1, x_2)\|^2\, \mathrm{d}\boldsymbol{x} \\ &= \frac{1}{2} \int_{\mathbb{R}^2} \boldsymbol{E}(x_1, x_2) \cdot \boldsymbol{k}\left(\frac{x_1}{\eta}\right) \cdot \boldsymbol{E}(x_1, x_2)\, \mathrm{d}\boldsymbol{x} \end{aligned} \tag{2.6}$$

where $\mu(x_1)$ denotes the local shear modulus, $\mu(x_1) = \mu_I$ if $\eta^{-1} x_1 \in \chi$ and $\mu(x_1) = \mu_{II}$ if $\eta^{-1} x_1 \notin \chi$, and $\|\cdot\|$ is the Euclidean norm of the vector $\boldsymbol{E}$. The elastic stiffness $\boldsymbol{k}(y)$ is the periodic function with period 1 of the stretched variable $y = \eta^{-1} x_1$

$$\boldsymbol{k}(y) = \boldsymbol{I}_2 \times \begin{cases} \mu_I & \text{if } y \in \chi, \\ \mu_{II} & \text{if } y \notin \chi. \end{cases} \tag{2.7}$$

Here, $\boldsymbol{I}_2$ is the identity matrix of dimension 2. Since both $\mu_I$ and $\mu_{II}$ are strictly positive, the symmetric matrix $\boldsymbol{k}(y)$ is also strictly positive in the sense of quadratic forms,

$$\forall y, \quad \boldsymbol{k}(y) > 0. \tag{2.8}$$

## 2.2 Series solution for the microscopic strain

We now consider the limit of well-separated scales,

$$\eta \ll 1, \tag{2.9}$$

in which homogenization is applicable. We use formal two-scale expansions and seek the displacement $v(x_1, x_2)$ in (2.3) in the form of a series expansion having dependence on both the slow variable $\boldsymbol{x} = (x_1, x_2)$ and on the fast variable $y = \eta^{-1} x_1$,

$$v(\boldsymbol{x}) = \sum_{i=0}^{+\infty} \eta^i\, v_i(\boldsymbol{x}, \eta^{-1} x_1). \tag{2.10}$$

The 'coefficients' $v_i(\boldsymbol{x}, y)$ are assumed to be smooth functions of their arguments $\boldsymbol{x}$ and $y$, and can be sought as periodic functions of $y$ without loss of generality. In (2.10), we have introduced a single stretched variable $y = \eta^{-1} x_1$ as the laminate is periodic in the direction $\boldsymbol{i}_1$.

The leading-order displacement $v_0(\boldsymbol{x}, y)$ can be identified with the macroscopic displacement $u(\boldsymbol{x})$ and does not depend on $y$. The higher-order contributions $v_i(\boldsymbol{x}, y)$ for $i > 0$ are found [Le15] as the solutions of successive problems formulated on the unit cell $y \in [0, 1]$, using asymptotic homogenization. They are proportional to the successive gradients of the macroscopic strain $\boldsymbol{\varepsilon}(\boldsymbol{x})$: the solution derived in [Le15] can be written as

$$v(\boldsymbol{x}) = u(\boldsymbol{x}) + \eta\, \boldsymbol{\Psi}_0(\eta^{-1} x_1) \cdot \boldsymbol{\varepsilon}(\boldsymbol{x}) + \eta^2\, \boldsymbol{\Psi}_1(\eta^{-1} x_1) \cdot \nabla \boldsymbol{\varepsilon}(\boldsymbol{x}) + \eta^3\, \boldsymbol{\Psi}_2(\eta^{-1} x_1) \cdot \nabla^2 \boldsymbol{\varepsilon}(\boldsymbol{x}) + \mathcal{O}(\eta^4). \tag{2.11}$$

In our notation, the macroscopic (effective) strain $\boldsymbol{\varepsilon}$ and its gradients are presented as vectors that are defined in block notation as

$$\boldsymbol{\varepsilon}(\boldsymbol{x}) = \nabla u(\boldsymbol{x}) = \begin{pmatrix} \frac{\partial u}{\partial x_1}(\boldsymbol{x}) \\ \frac{\partial u}{\partial x_2}(\boldsymbol{x}) \end{pmatrix} \in \mathbb{R}^2, \qquad \nabla \boldsymbol{\varepsilon}(\boldsymbol{x}) = \begin{pmatrix} \frac{\partial \boldsymbol{\varepsilon}}{\partial x_1}(\boldsymbol{x}) \\ \frac{\partial \boldsymbol{\varepsilon}}{\partial x_2}(\boldsymbol{x}) \end{pmatrix} \in \mathbb{R}^4, \qquad \nabla^2 \boldsymbol{\varepsilon}(\boldsymbol{x}) = \begin{pmatrix} \frac{\partial \nabla \boldsymbol{\varepsilon}}{\partial x_1}(\boldsymbol{x}) \\ \frac{\partial \nabla \boldsymbol{\varepsilon}}{\partial x_2}(\boldsymbol{x}) \end{pmatrix} = \begin{pmatrix} \frac{\partial^2 \boldsymbol{\varepsilon}}{\partial x_1^2}(\boldsymbol{x}) \\ \frac{\partial^2 \boldsymbol{\varepsilon}}{\partial x_1 \partial x_2}(\boldsymbol{x}) \\ \frac{\partial^2 \boldsymbol{\varepsilon}}{\partial x_1 \partial x_2}(\boldsymbol{x}) \\ \frac{\partial^2 \boldsymbol{\varepsilon}}{\partial x_2^2}(\boldsymbol{x}) \end{pmatrix} \in \mathbb{R}^8. \tag{2.12}$$

The macroscopic strain *vector* $\boldsymbol{\varepsilon}(\boldsymbol{x}) \in \mathbb{R}^2$ is a condensed representation of the classical strain *tensor* $\tilde{\boldsymbol{\varepsilon}}(\boldsymbol{x})$ from continuum mechanics, and the latter can be expressed in terms of the former by a formula similar to (2.5). Note that some of the entries in the strain gradients are redundant, such as $(\nabla^2\boldsymbol{\varepsilon})_3 = \partial[(\nabla\boldsymbol{\varepsilon})_3]/\partial x_1 = \partial^2\varepsilon_1/(\partial x_1\,\partial x_2) = \partial[(\nabla\boldsymbol{\varepsilon})_1]/\partial x_2 = (\nabla^2\boldsymbol{\varepsilon})_5$.

The first relocalization vectors $\boldsymbol{\Psi}_0(y) \in \mathbb{R}^2$ and $\boldsymbol{\Psi}_1(y) \in \mathbb{R}^4$ appearing in the homogenization solution (2.11) are obtained in subsection 2.4.2 of [Le15] in explicit form as

$$
\begin{aligned}
\boldsymbol{\Psi}_0(y) &= \begin{pmatrix} \dfrac{(\mu_I-\mu_{II})}{(1-a)\,\mu_I + a\,\mu_{II}} \times \begin{cases} (1-a)\,(\frac{a}{2}-y) & \text{if } y\in\chi \\ a\,(y-\frac{1+a}{2}) & \text{if } y\notin\chi \end{cases} \\ 0 \end{pmatrix} \\
\boldsymbol{\Psi}_1(y) &= \begin{pmatrix} -\dfrac{(\mu_I-\mu_{II})}{(1-a)\,\mu_I + a\,\mu_{II}} \times \begin{cases} (1-a)\left(\frac{a y}{2}-\frac{y^2}{2}+\frac{a\,(1-2a)}{12}\right) & \text{if } y\in\chi \\ a\left(\frac{y^2}{2}-\frac{1+a}{2}y+\frac{2a^2+3a+1}{12}\right) & y\notin\chi \end{cases} \\ 0 \\ 0 \\ -(\mu_I-\mu_{II})\left(\dfrac{a\,(1-a)}{12}\left(\dfrac{a^2}{\mu_I}-\dfrac{(1-a)^2}{\mu_{II}}\right) + \begin{cases} \frac{(1-a)}{\mu_I}\left(\frac{y^2}{2}-\frac{a y}{2}\right) & \text{if } y\in\chi \\ \frac{a}{\mu_{II}}\left(\frac{1+a}{2}y-\frac{y^2}{2}-\frac{a}{2}\right) & y\notin\chi \end{cases}\right) \end{pmatrix}.
\end{aligned}
\tag{2.13}
$$

By design, the operators $\boldsymbol{\Psi}_i(y)$ are all $y$-periodic and their average is zero,

$$
\langle\boldsymbol{\Psi}_0\rangle_y = \boldsymbol{0}_2, \quad \langle\boldsymbol{\Psi}_1\rangle_y = \boldsymbol{0}_4, \quad \ldots \tag{2.14}
$$

where $\langle\cdot\rangle_y$ represents the average operator on a unit-cell

$$
\langle\cdot\rangle_y = \int_0^1 \cdot\,\mathrm{d}y. \tag{2.15}
$$

The microscopic strain $\boldsymbol{E}$ can be obtained by inserting the solution (2.11) for the microscopic displacement into (2.4). The result is of the form $\boldsymbol{E}(\boldsymbol{x}) = \boldsymbol{E}^\star(\boldsymbol{x}, \eta^{-1}x_1)$ where $\boldsymbol{E}^\star(\boldsymbol{x}, y)$ is the smooth function, defined in block-matrix notation as

$$
\begin{aligned}
\boldsymbol{E}^\star(\boldsymbol{x},y) &= \boldsymbol{F}_0(y)\cdot\boldsymbol{\varepsilon}(\boldsymbol{x}) + \eta\,\boldsymbol{F}_1(y)\cdot\nabla\boldsymbol{\varepsilon}(\boldsymbol{x}) + \eta^2\boldsymbol{F}_2(y)\cdot\nabla^2\boldsymbol{\varepsilon}(\boldsymbol{x}) + \mathcal{O}(\eta^3) \\
&= \begin{pmatrix} \boldsymbol{F}_0(y) & \boldsymbol{F}_1(y) & \boldsymbol{F}_2(y) \end{pmatrix}\cdot\begin{pmatrix} \boldsymbol{\varepsilon}(\boldsymbol{x}) \\ \eta\,\nabla\boldsymbol{\varepsilon}(\boldsymbol{x}) \\ \eta^2\,\nabla^2\boldsymbol{\varepsilon}(\boldsymbol{x}) \end{pmatrix} + \mathcal{O}(\eta^3).
\end{aligned}
\tag{2.16}
$$

The strain relocalization matrices $\boldsymbol{F}_0, \boldsymbol{F}_1, \boldsymbol{F}_2, \ldots$ have respective dimensions $2\times 2$, $2\times 4$, $2\times 8$, … and are given by

$$
\begin{aligned}
\boldsymbol{F}_0(y) &= \boldsymbol{I}_2 + \boldsymbol{i}_1\otimes\boldsymbol{\Psi}_0'(y) \\
\boldsymbol{F}_1(y) &= \boldsymbol{i}_1\otimes\begin{pmatrix}\boldsymbol{\Psi}_0(y) & \boldsymbol{0}_2\end{pmatrix} + \boldsymbol{i}_2\otimes\begin{pmatrix}\boldsymbol{0}_2 & \boldsymbol{\Psi}_0(y)\end{pmatrix} + \boldsymbol{i}_1\otimes\boldsymbol{\Psi}_1'(y) \\
\boldsymbol{F}_2(y) &= \boldsymbol{i}_1\otimes\begin{pmatrix}\boldsymbol{\Psi}_1(y) & \boldsymbol{0}_4\end{pmatrix} + \boldsymbol{i}_2\otimes\begin{pmatrix}\boldsymbol{0}_4 & \boldsymbol{\Psi}_1(y)\end{pmatrix} + \boldsymbol{i}_1\otimes\boldsymbol{\Psi}_2'(y),
\end{aligned}
\tag{2.17}
$$

where $\cdot'$ denotes derivation with respect to $y$ and $(\,\boldsymbol{s}\ \boldsymbol{t}\,)$ denotes the vector obtained by concatenating the vectors $\boldsymbol{s}$ and $\boldsymbol{t}$ in block-vector notation. In the right-hand side of (2.17), the terms in the first and second column come from differentiating (2.11) with respect to $\boldsymbol{x}$ and $\eta^{-1}x_1$ respectively.

In view of (2.14), the averages of the relocalization matrices are $\langle\boldsymbol{F}_0\rangle_y = \boldsymbol{I}_2$, $\langle\boldsymbol{F}_1\rangle_y = \boldsymbol{0}$, $\langle\boldsymbol{F}_2\rangle_y = \boldsymbol{0}$, etc. By (2.16), this ensures that the macroscopic strain coincides with the average of the microscopic strain with respect to the fast variable, $\boldsymbol{\varepsilon}(\boldsymbol{x}) = \langle\boldsymbol{E}^\star(\boldsymbol{x},y)\rangle_y$.

## 2.3 Series expansion of the energy

In the limit of well-separated scales $\eta\to 0$ and up to boundary terms which we can ignore since the laminate covers the entire plane $\mathbb{R}^2$, the strain energy $\Phi[\boldsymbol{E}]$ in (2.6) can be integrated sequentially, over the fast variable $y$ first and over the slow variable $\boldsymbol{x}$ next,

$$
\begin{aligned}
\Phi[\boldsymbol{E}] &= \frac{1}{2}\int_{\mathbb{R}^2}\boldsymbol{E}(\boldsymbol{x})\cdot\boldsymbol{k}\left(\frac{x_1}{\eta}\right)\cdot\boldsymbol{E}(\boldsymbol{x})\,\mathrm{d}\boldsymbol{x} \\
&= \frac{1}{2}\int_{\mathbb{R}^2}\left\langle\boldsymbol{E}^\star(\boldsymbol{x},y)\cdot\boldsymbol{k}(y)\cdot\boldsymbol{E}^\star(\boldsymbol{x},y)\right\rangle_y d\boldsymbol{x}.
\end{aligned}
\tag{2.18}
$$

This identity is at the heart of energy-based homogenization and is used, *e.g.*, in Equation [2.5] of [Le15] .

Inserting the series solution (2.16) into the right-hand side in equation above, we obtain the strain energy $\Phi^\star[\boldsymbol{\varepsilon}] = \Phi[\boldsymbol{E}]$ in the form of a functional of the macroscopic strain $\boldsymbol{\varepsilon}$,

$$
\Phi^\star[\boldsymbol{\varepsilon}] = \int_{\mathbb{R}^2}\frac{1}{2}\begin{pmatrix} \boldsymbol{\varepsilon}(\boldsymbol{x}) \\ \eta\,\nabla\boldsymbol{\varepsilon}(\boldsymbol{x}) \\ \eta^2\,\nabla^2\boldsymbol{\varepsilon}(\boldsymbol{x}) \end{pmatrix}\cdot\boldsymbol{t}\cdot\begin{pmatrix} \boldsymbol{\varepsilon}(\boldsymbol{x}) \\ \eta\,\nabla\boldsymbol{\varepsilon}(\boldsymbol{x}) \\ \eta^2\,\nabla^2\boldsymbol{\varepsilon}(\boldsymbol{x}) \end{pmatrix}\mathrm{d}\boldsymbol{x} + \mathcal{O}(\eta^3) \tag{2.19}
$$

where the matrix of homogenized elastic properties $\boldsymbol{t}$ is the symmetric matrix of dimensions $14\times 14$ defined by

$$\boldsymbol{t} = \Big\langle (\,\boldsymbol{F}_0(y)\;\;\boldsymbol{F}_1(y)\;\;\boldsymbol{F}_2(y)\,)^T\cdot\boldsymbol{k}(y)\cdot(\,\boldsymbol{F}_0(y)\;\;\boldsymbol{F}_1(y)\;\;\boldsymbol{F}_2(y)\,)\Big\rangle_y. \tag{2.20}$$

Note that the block matrix $(\,\boldsymbol{F}_0\;\boldsymbol{F}_1\;\boldsymbol{F}_2\,)$ is of size $2\times(2+4+8)=2\times 14$, and that both contractions take place over the 2 rows of this matrix. The quantity $\boldsymbol{t}$ can be expressed in terms of $\mu_I$, $\mu_{II}$ and $a$ using (2.13) and (2.17) but the resulting expressions are cumbersome and are not included here.

The symmetric matrix $\boldsymbol{t}$ is clearly positive in view of the positivity of $\boldsymbol{k}(y)$ in (2.8),

$$\boldsymbol{t}\geqslant 0, \tag{2.21}$$

and our goal will be to preserve this positivity in the final homogenized energy.

The laminate being mirror-symmetric in both the horizontal and vertical directions, the homogenized energy $\Phi^\star$ can only couple gradients of the strain whose orders are either both even, or both odd: these are the only energy terms whose sign is preserved by the mirror symmetries. This is a classical result known as centro-symmetry in the context of periodic homogenization, see Appendix B of [Bou96] and [ABB09, TB96]. An explicit calculation of $\boldsymbol{t}$ using a symbolic calculation language confirms that the blocks in the matrix $\boldsymbol{t}$ that couple even and odd orders in the strain gradients are zero, so that $\boldsymbol{t}$ can be rewritten in block-matrix notation as

$$\boldsymbol{t} = \begin{pmatrix} \boldsymbol{K} & \boldsymbol{0}_{2\times 4} & \boldsymbol{C} \\ \boldsymbol{0}_{4\times 2} & \boldsymbol{B} & \boldsymbol{0}_{4\times 8} \\ \boldsymbol{C}^T & \boldsymbol{0}_{8\times 4} & \boldsymbol{Z} \end{pmatrix}. \tag{2.22}$$

In view of (2.19), the diagonal terms $\boldsymbol{K}$, $\boldsymbol{B}$, $\boldsymbol{Z}$ are square matrices with respective dimensions $2\times 2$, $4\times 4$ and $8\times 8$ that give rise to energy contributions proportional to $\boldsymbol{\varepsilon}\otimes\boldsymbol{\varepsilon}$, $\eta^2\,\nabla\boldsymbol{\varepsilon}\otimes\nabla\boldsymbol{\varepsilon}$ and $\eta^4\,\nabla^2\boldsymbol{\varepsilon}\otimes\nabla^2\boldsymbol{\varepsilon}$, respectively, whereas the off-diagonal block $\boldsymbol{C}$ has dimensions $2\times 8$ and gives rise to an energy contribution proportional to $\eta^2\,\boldsymbol{\varepsilon}\otimes\nabla^2\boldsymbol{\varepsilon}$.

Equation (2.19) can be rewritten in expanded form as

$$\Phi^\star[\boldsymbol{\varepsilon}] = \int_{\mathbb{R}^2}\frac{1}{2}\left(\boldsymbol{\varepsilon}\cdot\boldsymbol{K}\cdot\boldsymbol{\varepsilon}+\eta^2\,(\nabla\boldsymbol{\varepsilon}\cdot\boldsymbol{B}\cdot\nabla\boldsymbol{\varepsilon}+2\,\boldsymbol{\varepsilon}\cdot\boldsymbol{C}\cdot\nabla^2\boldsymbol{\varepsilon})+\eta^4\,\nabla^2\boldsymbol{\varepsilon}\cdot\boldsymbol{Z}\cdot\nabla^2\boldsymbol{\varepsilon}\right)\mathrm{d}\boldsymbol{x}+\mathcal{O}(\eta^3). \tag{2.23}$$

The sub-blocks of $\boldsymbol{t}$ associated with the successive orders in $\eta$ are therefore arranged along lines parallel to the SW-NE diagonal, stacked form NW (order $\eta^0$) to SE (order $\eta^4$), using standard abbreviations for intercardinal directions:

- the contribution of order $\eta^0$ comes from the top-left block $\boldsymbol{K}$ which contains the results of leading-order (classical) homogenization,
- the diagonal $\boldsymbol{C}^T$–$\boldsymbol{B}$–$\boldsymbol{C}$ gives rise to energy contributions proportional to $\eta^2$,
- the lower-right block $\boldsymbol{Z}$ gives rise to energy contributions proportional to $\eta^4$—these contributions are negligible in view of the remainder $\mathcal{O}(\eta^3)$ in (2.23) but we keep it for a moment.

With help of (2.20), we obtain the following expressions of the blocks making up the matrix $\boldsymbol{t}$ in (2.20)

| order | diagonal blocks | off-diagonal blocks |
|---|---|---|
| $\eta^0$ | $\boldsymbol{K} = \langle\boldsymbol{F}_0^T(y)\cdot\boldsymbol{k}(y)\cdot\boldsymbol{F}_0(y)\rangle_y$ | |
| $\eta^2$ | $\boldsymbol{B} = \langle\boldsymbol{F}_1^T(y)\cdot\boldsymbol{k}(y)\cdot\boldsymbol{F}_1(y)\rangle_y$ | $\boldsymbol{C} = \langle\boldsymbol{F}_0^T(y)\cdot\boldsymbol{k}(y)\cdot\boldsymbol{F}_2(y)\rangle_y$ |
| $\eta^4$ | $\boldsymbol{Z} = \langle\boldsymbol{F}_2^T(y)\cdot\boldsymbol{k}(y)\cdot\boldsymbol{F}_2(y)\rangle_y$ | |

(2.24)

By (2.8) or (2.21), the symmetric blocks $\boldsymbol{K}$ and $\boldsymbol{B}$ appearing on the diagonal of $\boldsymbol{t}$ are positive in the sense of quadratic forms,

$$\begin{array}{rcl}\boldsymbol{K} & > & 0\\ \boldsymbol{B} & \geqslant & 0.\end{array} \tag{2.25}$$

The negligible diagonal block $\boldsymbol{Z}$ is positive as well but this property will not be used. The strict positivity of $\boldsymbol{K}$ in (2.25) comes from the fact that $\boldsymbol{k}(y)>0$ and $\det\boldsymbol{F}_0(y) = [\,(1-a)\,\mu_I + a\,\mu_{II}]^{-1}\times\begin{cases}\mu_{II} & \text{if } y\in\chi\\ \mu_I & \text{if } y\notin\chi\end{cases}>0$, implying that $\boldsymbol{F}_0$ is full-rank.

**Remark 2.1.** The blocks $\boldsymbol{C}$ and $\boldsymbol{Z}$ are expressed in (2.24) in terms of $\boldsymbol{F}_2$, which is itself given in $(2.17)_3$ in terms of a high-order relocalization vector $\boldsymbol{\Psi}_2(y)$. The expression of $\boldsymbol{\Psi}_2(y)$ is not included in Equation (2.13) and can only be obtained by pushing homogenization to a higher order which leads to cumbersome calculations. There is in fact no need to compute $\boldsymbol{\Psi}_2(y)$:

- By the same argument as used at the end of Section C.3 in the Appendix of [AL23], one can establish the identity $\langle\boldsymbol{F}_0^T(y)\cdot\boldsymbol{k}(y)\cdot(\boldsymbol{i}_1\otimes\boldsymbol{\Psi}_2'(y))\rangle_y=\boldsymbol{0}$ which is a property of the leading-order solution $\boldsymbol{F}_0$, and holds irrespective of the expression of $\boldsymbol{\Psi}_2(y)$. This shows the correct value of $\boldsymbol{C}$ can be obtained by setting simply $\boldsymbol{\Psi}_2(y)=\boldsymbol{0}$.
- Setting $\boldsymbol{\Psi}_2(y)=\boldsymbol{0}$ does affect the value of $\boldsymbol{Z}$, but this has no consequence. Indeed, (*i*) $\boldsymbol{Z}$ gives rise to negligible energy contributions, of order $\eta^4$, and (*ii*) the only benefit of $\boldsymbol{Z}$ is to preserve the positivity of $\boldsymbol{t}$, see Section 2.5 below, but this positivity follows from (2.20), irrespective of the exact expression of $\boldsymbol{\Psi}_2(y)$. Setting $\boldsymbol{\Psi}_2(y)=\boldsymbol{0}$ is therefore acceptable.

## 2.4 Naïve truncation of the energy

In this section, we present the standard truncation procedure used in the literature on high-order homogenization, including in [Le15] among a number of other works, and point out its deficiencies.

The most obvious way of truncating the energy in (2.23) is to drop the negligible term of order $\eta^4$, which yields an approximation $\Phi^\dagger[\boldsymbol{\varepsilon}] = \Phi^\star[\boldsymbol{\varepsilon}] + \mathcal{O}(\eta^3)$ consistent with second-order homogenization in the form

$$\Phi^\dagger[\boldsymbol{\varepsilon}] = \int_{\mathbb{R}^2} \frac{1}{2} (\boldsymbol{\varepsilon}\cdot\boldsymbol{K}\cdot\boldsymbol{\varepsilon} + \eta^2 (\nabla\boldsymbol{\varepsilon}\cdot\boldsymbol{B}\cdot\nabla\boldsymbol{\varepsilon} + 2\,\boldsymbol{\varepsilon}\cdot\boldsymbol{C}\cdot\nabla^2\boldsymbol{\varepsilon}))\,\mathrm{d}\boldsymbol{x}. \tag{2.26}$$

This truncation amounts to trim off the blocks appearing below the SW-NE diagonal in the expression of $\boldsymbol{t}$ in (2.22), that is to replace $\boldsymbol{t}$ with

$$\boldsymbol{t}^\dagger = \begin{pmatrix} \boldsymbol{K} & \boldsymbol{0} & \boldsymbol{C} \\ \boldsymbol{0} & \boldsymbol{B} & \boldsymbol{0} \\ \boldsymbol{C}^T & \boldsymbol{0} & \boldsymbol{0} \end{pmatrix} \tag{2.27}$$

Next the $\boldsymbol{C}$-terms, proportional to $\nabla^2\boldsymbol{\varepsilon}$, are integrated by parts and merged with the $\boldsymbol{B}$-terms. This yields another approximation $\Phi^\ddagger[\boldsymbol{\varepsilon}]$ of the energy consistent with the previous one up to boundary terms (which we ignore since the laminate is infinite in all directions),

$$\begin{aligned}\Phi^\ddagger[\boldsymbol{\varepsilon}] &= \int_{\mathbb{R}^2} \left(\frac{1}{2}\boldsymbol{\varepsilon}(\boldsymbol{x})\cdot\boldsymbol{K}\cdot\boldsymbol{\varepsilon}(\boldsymbol{x}) + \frac{\eta^2}{2}\nabla\boldsymbol{\varepsilon}(\boldsymbol{x})\cdot\boldsymbol{B}^\ddagger\cdot\nabla\boldsymbol{\varepsilon}(\boldsymbol{x})\right)\mathrm{d}\boldsymbol{x} \\ &= \int_{\mathbb{R}^2}\frac{1}{2}\begin{pmatrix}\boldsymbol{\varepsilon}(\boldsymbol{x}) \\ \eta\,\nabla\boldsymbol{\varepsilon}(\boldsymbol{x}) \\ \eta^2\,\nabla^2\boldsymbol{\varepsilon}(\boldsymbol{x})\end{pmatrix}\cdot\begin{pmatrix}\boldsymbol{K} & \boldsymbol{0} & \boldsymbol{0} \\ \boldsymbol{0} & \boldsymbol{B}^\ddagger & \boldsymbol{0} \\ \boldsymbol{0} & \boldsymbol{0} & \boldsymbol{0}\end{pmatrix}\cdot\begin{pmatrix}\boldsymbol{\varepsilon}(\boldsymbol{x}) \\ \eta\,\nabla\boldsymbol{\varepsilon}(\boldsymbol{x}) \\ \eta^2\,\nabla^2\boldsymbol{\varepsilon}(\boldsymbol{x})\end{pmatrix}\mathrm{d}\boldsymbol{x}.\end{aligned} \tag{2.28}$$

where the gradient properties are now captured by the symmetric matrix

$$\boldsymbol{B}^\ddagger = \boldsymbol{B} - (\tilde{\boldsymbol{C}} + \tilde{\boldsymbol{C}}^T). \tag{2.29}$$

Here, $\tilde{\boldsymbol{C}}$ is the $4\times4$ matrix obtained by splitting the matrix $\boldsymbol{C} = (\,\boldsymbol{C}_1\ \ \boldsymbol{C}_2\,)$ of dimensions $2\times8$ into two blocks $\boldsymbol{C}_1$ and $\boldsymbol{C}_2$ having dimensions $2\times4$ each, and rearranging them as

$$\tilde{\boldsymbol{C}} = \begin{pmatrix}\boldsymbol{C}_1 \\ \boldsymbol{C}_2\end{pmatrix}.$$

The new tensor $\boldsymbol{B}^\ddagger$ capturing the gradient effect is not positive. Indeed, its *diagonal* component $B^\ddagger_{44}$, calculated with the help of Equation [2.89] in [Le15], is strictly negative,

$$B^\ddagger_{44} = -\frac{a^2\,(1-a)^2\,(\mu_I-\mu_{II})^2\,((1-a)\,\mu_I + a\,\mu_{II})}{12\,\mu_I\,\mu_{II}} < 0. \tag{2.30}$$

With a non-positive matrix of high-order homogenized moduli $\boldsymbol{B}^\ddagger$, the energy $\Phi^\ddagger[\boldsymbol{\varepsilon}]$ is entirely useless. Indeed, observe that the mode $u_-(x_1,x_2) = \sin(\eta^{-1}x_2)$ has strain $\boldsymbol{\varepsilon}_- = (\,0\ \ \eta^{-1}\cos(\eta^{-1}x_2)\,)$, strain gradient $\nabla\boldsymbol{\varepsilon}_- = -(\,0\ \ 0\ \ 0\ \ \eta^{-2}\sin(\eta^{-1}x_2)\,)$, and contributes to the energy density through the gradient term as $\frac{\eta^2}{2}B^\ddagger_{44}\,(\nabla\boldsymbol{\varepsilon}_-)_4^2 = \frac{\eta^{-2}}{2}B^\ddagger_{44}\sin^2(\eta^{-1}x_2)$, which on average over $x_2$ and for small $\eta$ is a very large, negative contribution. An arbitrary equilibrium solution $u$ that makes the energy $\Phi^\ddagger[\boldsymbol{\varepsilon}]$ stationary can be combined with a perturbation along the mode $u_-$ with a small amplitude $\mathcal{O}(\eta^{1/2}) \ll 1$ to make the energy as large and negative as one wishes. This shows that the energy $\Phi^\ddagger[\boldsymbol{\varepsilon}]$ is not bounded from below in any neighborhood of any particular solution. In particular, no equilibrium is stable.

In the remainder of Section 2, we explore an alternative truncation method that preserves the positivity of $\boldsymbol{t}$ and requires no integration by parts.

## 2.5 Over-conservative truncation

In recent work [KA24], we have proposed to skip the truncation altogether as a way to preserve positivity: an energy functional similar to that appearing in (2.23) has been used, including the high-order $\boldsymbol{Z}$ terms,

$$\int_{\mathbb{R}^2}\frac{1}{2}(\boldsymbol{\varepsilon}\cdot\boldsymbol{K}\cdot\boldsymbol{\varepsilon} + \eta^2(\nabla\boldsymbol{\varepsilon}\cdot\boldsymbol{B}\cdot\nabla\boldsymbol{\varepsilon} + 2\,\boldsymbol{\varepsilon}\cdot\boldsymbol{C}\cdot\nabla^2\boldsymbol{\varepsilon}) + \eta^4\,\nabla^2\boldsymbol{\varepsilon}\cdot\boldsymbol{Z}\cdot\nabla^2\boldsymbol{\varepsilon})\,\mathrm{d}\boldsymbol{x}. \tag{2.31}$$

The work [KA24] was in fact concerned with nonlinear dimension reduction and is being rephrased here in the language of linear homogenization. The positivity of the $\boldsymbol{t}$ matrix, see (2.21), warrants that the homogenized energy above does not suffer from the deficiencies discussed in the previous section. The energy functional could successfully be used in numerical simulations to generate self-contained, coarse-grained predictions having higher-order accuracy [KA24].

This approach requires to explicitly calculate the 'negligible' $\boldsymbol{Z}$ block for the sole purpose of preserving positivity. Calculating $\boldsymbol{Z}$ is tedious, as shown by the (incomplete) set of expressions given in the Appendix of [KA24]. This work concerned a ribbon, in dimension $d = 1$, and the size of the square matrix $\boldsymbol{Z}$ grows quite quickly with the dimension $d$, making the calculation of $\boldsymbol{Z}$ even less desirable.

## 2.6 Minimal truncation preserving positivity

We proceed to present the new truncation method which preserves positivity without computing the $\boldsymbol{Z}$-block. It only uses the asymptotically meaningful blocks $\boldsymbol{K}$, $\boldsymbol{B}$ and $\boldsymbol{C}$ as input, *i.e.*, it starts from the truncated matrix of elastic moduli,

$$\boldsymbol{t} = \begin{pmatrix} \boldsymbol{K} & \boldsymbol{0}_{2\times 4} & \boldsymbol{C} \\ \boldsymbol{0}_{4\times 2} & \boldsymbol{B} & \cdots \\ \boldsymbol{C}^T & \cdots & \cdots \end{pmatrix}. \tag{2.32}$$

where the dots denote blocks associated with higher-order effects.

Instead of integrating the $\boldsymbol{\varepsilon} \otimes \nabla^2 \boldsymbol{\varepsilon}$ terms by parts, we incorporate them into the leading-order terms: up to terms of order $\eta^3$ denoted by the ellipses, we have

$$\begin{aligned} \Phi^\star[\boldsymbol{\varepsilon}] &= \int_{\mathbb{R}^2} \frac{1}{2} \left(\boldsymbol{\varepsilon} \cdot \boldsymbol{K} \cdot \boldsymbol{\varepsilon} + \eta^2 \left(\nabla \boldsymbol{\varepsilon} \cdot \boldsymbol{B} \cdot \nabla \boldsymbol{\varepsilon} + 2 \boldsymbol{\varepsilon} \cdot \boldsymbol{C} \cdot \nabla^2 \boldsymbol{\varepsilon}\right)\right) \mathrm{d}\boldsymbol{x} + \cdots \\ &= \int_{\mathbb{R}^2} \frac{1}{2} \left(\left\{\boldsymbol{\varepsilon} + \eta^2 \boldsymbol{K}^{-1} \cdot \boldsymbol{C} \cdot \nabla^2 \boldsymbol{\varepsilon}\right\} \cdot \boldsymbol{K} \cdot \left\{\boldsymbol{\varepsilon} + \eta^2 \boldsymbol{K}^{-1} \cdot \boldsymbol{C} \cdot \nabla^2 \boldsymbol{\varepsilon}\right\} + \eta^2 \nabla \boldsymbol{\varepsilon} \cdot \boldsymbol{B} \cdot \nabla \boldsymbol{\varepsilon}\right) \mathrm{d}\boldsymbol{x} + \cdots \end{aligned} \tag{2.33}$$

The right-hand sides on the first and second line differ only by a negligible term, $\eta^4 \left(\boldsymbol{C} \cdot \nabla^2 \boldsymbol{\varepsilon}\right) \cdot \boldsymbol{K}^{-1} \cdot \left(\boldsymbol{C} \cdot \nabla^2 \boldsymbol{\varepsilon}\right)$. We have just identified another asymptotically correct approximation of the energy at this order, $\Phi_+[\boldsymbol{\varepsilon}] = \Phi^\star[\boldsymbol{\varepsilon}] + \mathcal{O}(\eta^3)$, in the form

$$\Phi_+[\boldsymbol{\varepsilon}] = \int_{\mathbb{R}^2} \begin{pmatrix} \frac{1}{2} \left\{\boldsymbol{\varepsilon} + \eta^2 \boldsymbol{K}^{-1} \cdot \boldsymbol{C} \cdot \nabla^2 \boldsymbol{\varepsilon}\right\} \cdot \boldsymbol{K} \cdot \left\{\boldsymbol{\varepsilon} + \eta^2 \boldsymbol{K}^{-1} \cdot \boldsymbol{C} \cdot \nabla^2 \boldsymbol{\varepsilon}\right\} \\ + \frac{\eta^2}{2} \nabla \boldsymbol{\varepsilon} \cdot \boldsymbol{B} \cdot \nabla \boldsymbol{\varepsilon} \end{pmatrix} \mathrm{d}\boldsymbol{x}. \tag{2.34}$$

In addition to being asymptotically correct, this energy is positive thanks to the positivity of the symmetric matrices $\boldsymbol{K} > 0$ and $\boldsymbol{B} \geqslant 0$, see (2.25). The strict positivity of $\boldsymbol{K}$ also ensures that it can be inverted.

The energy density in (2.34) is made up of two terms:

- On the first line, the leading-order homogenized elastic coefficients are combined with a *generalized strain* appearing in curly braces, which is the classical strain $\boldsymbol{\varepsilon}$ at leading order corrected by a second-gradient term $\eta^2 \boldsymbol{K}^{-1} \cdot \boldsymbol{C} \cdot \nabla^2 \boldsymbol{\varepsilon}$;
- On the second line, the positive matrix of higher-order moduli $\boldsymbol{B}$ is combined with the standard strain gradient.

The key property of $\Phi_+$ in (2.34) is that the $\boldsymbol{C}$ block, which has no special positivity property—it is not even a square matrix—, enters in the definition of the generalized strain (inside the arguments of the quadratic forms) but *not* in the definition of the elastic moduli (in the quadratic forms themselves). By contrast, the writing in (2.26) suggests that the entries in $\boldsymbol{C}$ can be interpreted as homogenized elastic moduli, which is misleading.

The homogenized energy $\Phi_+[\boldsymbol{\varepsilon}]$ is therefore both asymptotically correct and simple to derive.

## 2.7 Notation recap

In Table 2.1, we recapitulate the symbols used for the energy at the successive steps of homogenization. The microscopic energy we started from is denoted as $\Phi[\boldsymbol{E}]$, where $\boldsymbol{E}$ is the microscopic strain. Using homogenization, we have been able to eliminate the macroscopic variables in favor of the macroscopic strain $\boldsymbol{\varepsilon}$ and derive a functional $\Phi^\star[\boldsymbol{\varepsilon}]$. Next, we have discussed two strategies for truncating $\Phi^\star$. In existing work, all terms that are negligible are removed, yielding a functional $\Phi^\dagger[\boldsymbol{\varepsilon}]$ which is subsequently integrated by parts into $\Phi^\ddagger[\boldsymbol{\varepsilon}]$. The latter, however, is non-positive. We proposed an alternative truncation yielding a *non-negative* energy functional $\Phi_+ \geqslant 0$. All these energy functional agree up to an error of order $\eta^{p+1}$, where $\eta \ll 1$ is the scale separation parameter and $p$ is the homogenization order—we use $p = 2$ throughout in this paper.

| | |
|---|---|
| microscopic model | $\Phi[\boldsymbol{E}]$ |
| output of homogenization | $\Phi^\star[\boldsymbol{\varepsilon}]$ |
| naïve truncation | $\Phi^\dagger[\boldsymbol{\varepsilon}]$, $\Phi^\ddagger[\boldsymbol{\varepsilon}]$ |
| positive truncation | $\Phi_+[\boldsymbol{\varepsilon}]$ |

**Table 2.1.** Main steps in homogenization procedure and symbols used for the different types of energy functionals.

# 3 A simple lattice: the triangular truss

In this section, we show that the truncation procedure proposed in Section 2.6 can be applied to *discrete* elastic structures as well. We start with the triangular truss in dimension 2, shown in Figure 3.1. In a simple lattice such as the triangular truss, the nodal displacement is expressed directly in terms of the macroscopic displacement and there is no need to introduce a microscopic shift. This makes homogenization straightforward. Non-positivity issues do arise in the homogenized energy of simple lattices, however, and they can be fixed by the same strategy as that presented in Section 2.6 for the laminate, as we show. The case of a complex lattice is addressed in the next section.

This section and the next one are based on the high-order homogenization method for elastic lattices presented in [YAL24]. For a detailed presentation, the reader is referred to this paper.

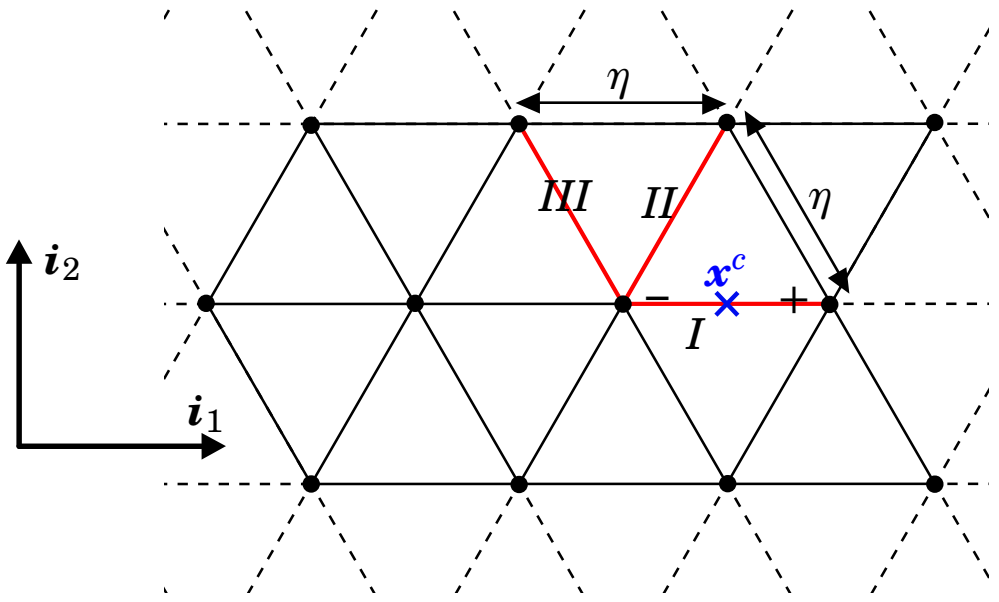

**Figure 3.1.** A 2D infinite periodic triangular lattice, composed of identical springs with spring constant $k$. A unit cell is shown in red, and contains three springs of types $\varphi \in \{I,II,III\}$. The center of a particular spring analyzed in the text is denoted as $\boldsymbol{x}^c$ and its endpoints as $\pm$.

## 3.1 Geometry and elastic energy

The triangular lattice is shown in Figure 3.1. It is obtained by repeating periodically the unit-cell, shown in red in the figure, along the generating vectors $\eta\,\boldsymbol{i}_1$ and $\eta\,\boldsymbol{i}_2' = \eta\,\left(\boldsymbol{i}_1 - \sqrt{3}\,\boldsymbol{i}_2\right)/2$, where $\eta \ll 1$ is the length of the truss elements. There are three families of springs with different orientations in space, having all the same rest length $\eta$ and spring constant $k$. The families are labelled by an index $\varphi \in \{I, II, III\}$, with $\varphi = I$ for the springs aligned with the $\boldsymbol{i}_1$ direction.

In this simple lattice, homogenization is carried out by writing the nodal displacement $\boldsymbol{v}_\beta$ of a node labelled $\beta$ having undeformed position $\boldsymbol{x}_\beta$ in terms of a macroscopic displacement $\boldsymbol{u}: \mathbb{R}^2 \to \mathbb{R}^2$ as

$$\boldsymbol{v}_\beta = \boldsymbol{u}(\boldsymbol{x}_\beta). \tag{3.1}$$

Next, the linearized change of length of the different springs is calculated. Consider the particular spring of type $I$ having midpoint (center) $\boldsymbol{x}^c$ shown in Figure 3.1, for instance. We denote its endpoints as $\beta = -$ and $\beta = +$, having undeformed position $\boldsymbol{X}_\pm = \boldsymbol{x}^c \pm \frac{\eta}{2}\boldsymbol{i}_1$, respectively. The linearized change of length of the spring can be written as

$$\begin{aligned}
(\boldsymbol{v}_+ - \boldsymbol{v}_-)\cdot\boldsymbol{i}_1 &= \left(\boldsymbol{u}\left(\boldsymbol{x}^c + \frac{\eta}{2}\boldsymbol{i}_1\right) - \boldsymbol{u}\left(\boldsymbol{x}^c - \frac{\eta}{2}\boldsymbol{i}_1\right)\right)\cdot\boldsymbol{i}_1 \\
&= \eta\left(\frac{\partial u_1}{\partial x_1}(\boldsymbol{x}^c) + \frac{\eta^2}{3!\,2^2}\frac{\partial^3 u_1}{\partial x_1^3}(\boldsymbol{x}^c) + \mathcal{O}(\eta^4)\right) \\
&= \eta\left(\varepsilon_{11}(\boldsymbol{x}^c) + \frac{\eta^2}{3!\,2^2}\frac{\partial^2 \varepsilon_{11}}{\partial x_1^2}(\boldsymbol{x}^c) + \mathcal{O}(\eta^4)\right) \\
&= \eta\left([\boldsymbol{i}_1\otimes\boldsymbol{i}_1] : \boldsymbol{\varepsilon}(\boldsymbol{x}^c) + \eta^2\left[\frac{1}{3!\,2^2}\boldsymbol{i}_1\otimes\boldsymbol{i}_1\otimes\boldsymbol{i}_1\otimes\boldsymbol{i}_1\right] :: \nabla^2\boldsymbol{\varepsilon}(\boldsymbol{x}^c) + \mathcal{O}(\eta^4)\right),
\end{aligned} \tag{3.2}$$

where the symbols $:$ and $::$ represent multiple contraction, see Appendix A for a proper definition, and $\boldsymbol{\varepsilon}(\boldsymbol{x})$ denotes the macroscopic strain,

$$\boldsymbol{\varepsilon}(\boldsymbol{x}) = \frac{1}{2}\left(\nabla\boldsymbol{u}(\boldsymbol{x}) + \nabla\boldsymbol{u}^T(\boldsymbol{x})\right). \tag{3.3}$$

The strain $\boldsymbol{\varepsilon}(\boldsymbol{x})$ is now a symmetric, $2\times 2$ matrix as we no longer use the compact, *vector* form of the strain used earlier in (2.12), that was relevant to antiplane elasticity. The successive strain gradients $\nabla^k\boldsymbol{\varepsilon}$ are now tensors of order $k+2$ and dimensions $2\times\ldots\times 2$ and are no longer represented in compact, vector form either.

In view of (3.2), we introduce the *continuous* function

$$E_I(\boldsymbol{x}) = [\boldsymbol{i}_1\otimes\boldsymbol{i}_1] : \boldsymbol{\varepsilon}(\boldsymbol{x}) + \eta^2\left[\frac{1}{3!\,2^2}\boldsymbol{i}_1\otimes\boldsymbol{i}_1\otimes\boldsymbol{i}_1\otimes\boldsymbol{i}_1\right] :: \nabla^2\boldsymbol{\varepsilon}(\boldsymbol{x}) + \mathcal{O}(\eta^4), \tag{3.4}$$

which yields the change of length in a spring of type $I$ as $\eta\,E_I(\boldsymbol{x}^c)$, where $\boldsymbol{x}^c$ is the spring midpoint position in undeformed configuration. The dependence of $E_I(\boldsymbol{x})$ on $\eta$ is implicit in our notation.

The argument is repeated with the two other families of springs and the quantities $E_I(\boldsymbol{x})$, $E_{II}(\boldsymbol{x})$ and $E_{III}(\boldsymbol{x})$ are collected in a vector called *microscopic strain*,

$$\boldsymbol{E}(\boldsymbol{x}) = \begin{pmatrix} E_I(\boldsymbol{x}) \\ E_{II}(\boldsymbol{x}) \\ E_{III}(\boldsymbol{x}) \end{pmatrix} \in \mathbb{R}^3. \tag{3.5}$$

Using (3.2) and similar formulas for the other types of springs, we can express $\boldsymbol{E}(\boldsymbol{x})$ in terms of the macroscopic strain $\boldsymbol{\varepsilon}$ as

$$\boldsymbol{E}(\boldsymbol{x}) = \boldsymbol{F}_0 : \boldsymbol{\varepsilon}(\boldsymbol{x}) + \eta^2\boldsymbol{F}_2 :: \nabla^2\boldsymbol{\varepsilon}(\boldsymbol{x}) + \mathcal{O}(\eta^3). \tag{3.6}$$

where the *constant* tensors $\boldsymbol{F}_0$ and $\boldsymbol{F}_2$, with respective dimensions $3\times2\times2$ and $3\times2\times2\times2\times2$ are obtained by assembling the coefficients appearing in square brackets in Equation (3.2) for type-$I$ springs and in the corresponding equations for the other types of springs.

Denoting as $i\in\varphi$ the index of a generic spring of type $\varphi$ and as $\boldsymbol{x}_i^{\mathrm{c}}$ its midpoint, we can write the energy of the discrete lattice as a sum of all springs, grouped by family,

$$\Phi = \sum_{\varphi\in\{I,II,III\}}\sum_{i\in\varphi}\frac{1}{2}k\left(\eta E_\varphi(\boldsymbol{x}_i^{\mathrm{c}})\right)^2. \tag{3.7}$$

Using the Euler–macLaurin formula [KC02], and up to boundary terms which we ignore as the domain is infinite, we can rewrite each sum (involving a particular type of links $\varphi$) as an integral

$$\begin{aligned}\Phi &= \sum_{\varphi\in\{I,II,III\}}\int_{\mathbb{R}^2}\frac{1}{2}\rho\,\eta^2 k\,E_\varphi^2(\boldsymbol{x})\,\mathrm{d}\boldsymbol{x}\\ &= \int_{\mathbb{R}^2}\frac{1}{2}\rho\,\eta^2 k\,\|\boldsymbol{E}(\boldsymbol{x})\|^2\,\mathrm{d}\boldsymbol{x}.\end{aligned} \tag{3.8}$$

Here, $\rho$ represents the density per unit area of any particular type of springs,

$$\rho = \frac{2}{\sqrt{3}\,\eta^2}. \tag{3.9}$$

The strain energy (3.8) can be rewritten in a way similar to (2.6–2.7) as

$$\Phi[\boldsymbol{E}] = \int_{\mathbb{R}^2}\frac{1}{2}\boldsymbol{E}(\boldsymbol{x})\cdot\boldsymbol{k}\cdot\boldsymbol{E}(\boldsymbol{x})\,\mathrm{d}\boldsymbol{x}, \tag{3.10}$$

where $\boldsymbol{k}$ is the stiffness matrix

$$\boldsymbol{k} = \frac{2k}{\sqrt{3}}\boldsymbol{I}_3. \tag{3.11}$$

Equation (3.10) follows from Equations [3.38,3.39,5.1] in [YAL24].

## 3.2 Series solution for the microscopic strain

As the lattice is simple, there are neither a microscopic shift nor a fast variable $y$, and the expansion (3.6) already yields the microscopic strain $\boldsymbol{E}$ in terms of the macroscopic strain $\boldsymbol{\varepsilon}$ and its gradients, like Equation (2.16) for the laminate.

In the companion notebook homogenize-triangular-truss.nb provided as supplementary material [TAL24b] to the present paper, we use the shoal library [Aud23], that carries out automatic second-order homogenization of elastic lattices based on the symbolic calculation language Wolfram Mathematica [Wol21]. The library is presented in detail in [AL23]. The geometrical and elastic features of the lattice are given to the library, which calculates the various quantities related to its homogenization, including the strain localization tensors $\boldsymbol{F}_0$ and $\boldsymbol{F}_2$: their detailed expressions are available in the companion Mathematica notebook.

## 3.3 Series expansion of the energy

By inserting the expansion of the strain $\boldsymbol{E}(\boldsymbol{x})$ in (3.6) into the continuous energy $\Phi[\boldsymbol{E}]$ in (3.10), we can define an effective energy $\Phi^\star[\boldsymbol{\varepsilon}]$ for the triangular lattice as

$$\Phi^\star[\boldsymbol{\varepsilon}] = \int_{\mathbb{R}^2}\frac{1}{2}\boldsymbol{\varepsilon}(\boldsymbol{x}):\boldsymbol{K}:\boldsymbol{\varepsilon}(\boldsymbol{x}) + \eta^2\,\boldsymbol{\varepsilon}(\boldsymbol{x}):\boldsymbol{C}::\nabla^2\boldsymbol{\varepsilon}(\boldsymbol{x})\,\mathrm{d}\boldsymbol{x} + \mathcal{O}(\eta^3) \tag{3.12}$$

where $\boldsymbol{K}$ and $\boldsymbol{C}$ are the fourth and sixth-tensors, respectively, given by

$$\begin{aligned}K_{ijkl} &= (F_0)_{pij}\,k_{pq}\,(F_0)_{qkl}\\ C_{ijklmn} &= (F_0)_{pij}\,k_{pq}\,(F_2)_{qklmn}.\end{aligned} \tag{3.13}$$

Here, we follow Einstein's summation convention, whereby a repeated index implies an implicit sum on this index. These expressions of $\boldsymbol{K}$ and $\boldsymbol{C}$ are similar to those derived for the laminate in (2.24).

The tensors $\boldsymbol{K}$ and $\boldsymbol{C}$ are directly output by the shoal library in symbolic form as the library already implements the formulas (3.13). With the help of the symbolic calculation language Wolfram Mathematica, we have rewritten their action on $\boldsymbol{\varepsilon}$ and $\nabla^2\boldsymbol{\varepsilon}$ in terms of invariants that reflect the symmetries of the triangular lattice. The details of the reduction are available from the companion notebook and the result is

$$\boldsymbol{K}:\boldsymbol{\varepsilon}(\boldsymbol{x}) = (\lambda+\mu)\,\boldsymbol{I}_2\,\mathrm{tr}\,\boldsymbol{\varepsilon} + 2\mu\,\boldsymbol{\varepsilon}^D, \tag{3.14}$$

where $\boldsymbol{\varepsilon}^D = \boldsymbol{\varepsilon} - \boldsymbol{I}_2\,\mathrm{tr}\,\boldsymbol{\varepsilon}/2$ denotes the deviatoric part of $\boldsymbol{\varepsilon}$ in dimension 2 and $\lambda$ and $\mu$ are the Lamé parameters predicted by leading-order homogenization,

$$\lambda = \mu = \frac{\sqrt{3}\,k}{4} > 0. \tag{3.15}$$

The expression of $\boldsymbol{K}$ in (3.14) is consistent with classical homogenization results from previous work, see [VP12, S18c] among others.

Equation (3.14) shows that the quadratic form $\boldsymbol{K}$ acting on symmetric tensors $\boldsymbol{\varepsilon}$ is strictly positive,

$$\boldsymbol{K} > 0. \tag{3.16}$$

The tensor $\boldsymbol{C}$ in (3.13) is reduced similarly using a symbolic calculation language and we obtain

$$\boldsymbol{C} :: \nabla^2 \boldsymbol{\varepsilon}(\boldsymbol{x}) = \frac{k}{32\sqrt{3}} \left( \nabla^2 \mathrm{tr}\, \boldsymbol{\varepsilon} + \Delta \boldsymbol{\varepsilon} + \frac{\boldsymbol{I}_2}{2} \mathscr{D} :: \nabla^2 \boldsymbol{\varepsilon} + \left[ \frac{\mathscr{J} : \nabla^2 \boldsymbol{\varepsilon} : \mathscr{J}^{T[3,1,2]}}{2} \right]^{\mathrm{sym}} \right), \tag{3.17}$$

where $\Delta \boldsymbol{\varepsilon} = \nabla^2 \boldsymbol{\varepsilon} : \boldsymbol{I}_2$ denotes the Laplacian of the strain,

$$\mathscr{J} = \boldsymbol{i}_1 \otimes (-\boldsymbol{i}_1 \otimes \boldsymbol{i}_1 + \boldsymbol{i}_2 \otimes \boldsymbol{i}_2) + \boldsymbol{i}_2 \otimes (\boldsymbol{i}_1 \otimes \boldsymbol{i}_2 + \boldsymbol{i}_2 \otimes \boldsymbol{i}_1) \tag{3.18}$$

is a rank-3 tensor that is invariant by the $D_6$ symmetry of the triangular lattice, $\mathscr{D}_{ijkl} = \mathscr{J}_{aij}\, \mathscr{J}_{akl}/2$ is the projector on the subspace of deviatoric strain, $\boldsymbol{s}^{T[3,1,2]}$ denotes the generalized transpose of a rank-3 tensor $\boldsymbol{s}$ such that $(s^{T[3,1,2]})_{jki} = s_{ijk}$, and $[\boldsymbol{i}]^{\mathrm{sym}} = (\boldsymbol{i} + \boldsymbol{i}^T)/2$ denotes the symmetric part of a $2 \times 2$ tensor.

**Remark 3.1.** Comparing (2.23) and (3.13) reveals that $\boldsymbol{B} = \boldsymbol{0}$ for the triangular lattice. This is a direct consequence of the fact that $\boldsymbol{F}_1 = \boldsymbol{0}$, as shown by comparing (2.16) and (3.6), see also the expression of $\boldsymbol{B}$ in (2.24).

**Remark 3.2.** Since we will be applying the new truncation method presented in Section 2.6 and will be bypassing the over-conservative one from Section 2.5 for this application (as well as for the forthcoming ones), we have not even derived the $\boldsymbol{Z}$ terms of order $\eta^4$ in the energy (3.12), unlike earlier in (2.23).

## 3.4 Naïve truncation of the energy

If we truncate and integrate by parts the term $\boldsymbol{\varepsilon}(\boldsymbol{x}) : \boldsymbol{C} :: \nabla^2 \boldsymbol{\varepsilon}(\boldsymbol{x})$ as we did earlier in (2.28), and as is commonly done in the literature, we get an energy $\Phi^{\ddagger}[\boldsymbol{\varepsilon}]$ in the form

$$\Phi^{\ddagger}[\boldsymbol{\varepsilon}] = \int_{\mathbb{R}^2} \frac{1}{2} \boldsymbol{\varepsilon}(\boldsymbol{x}) : \boldsymbol{K} : \boldsymbol{\varepsilon}(\boldsymbol{x}) - \frac{\eta^2}{2} \nabla \boldsymbol{\varepsilon}(\boldsymbol{x}) \therefore (2\boldsymbol{C}) \therefore \nabla \boldsymbol{\varepsilon}(\boldsymbol{x})\, \mathrm{d}\boldsymbol{x}. \tag{3.19}$$

This expression is similar to that in (2.28–2.29) for the laminate, but with $\boldsymbol{B}^{\ddagger} = -2\boldsymbol{C}$ since $\boldsymbol{B} = \boldsymbol{0}$. Note that the tensor $\boldsymbol{C}$ natively satisfies the major symmetries, $C_{ijklmn} = C_{lmnijk}$.

The quadratic form $\boldsymbol{B}^{\ddagger} = -2\boldsymbol{C}$ is calculated in symbolic form in the companion notebook, and all its eigenvalues are found to be negative, $\boldsymbol{B}^{\ddagger} \leqslant 0$, *i.e.*, $2\boldsymbol{C} \geqslant 0$. The energy $\Phi^{\ddagger}[\boldsymbol{\varepsilon}]$ is therefore non-positive and the naïve truncation fails again.

## 3.5 Truncation preserving positivity

As argued earlier in Section 2.6, incorporating the second-order term $\boldsymbol{\varepsilon}(\boldsymbol{x}) : \boldsymbol{C} :: \nabla^2 \boldsymbol{\varepsilon}(\boldsymbol{x})$ in (3.12) into the leading-order term $\boldsymbol{\varepsilon}(\boldsymbol{x}) : \boldsymbol{K} : \boldsymbol{\varepsilon}(\boldsymbol{x})$ preserves positivity. This yields an approximation of the energy that is exact to order $\eta^2$ included, $\Phi_+[\boldsymbol{\varepsilon}] = \Phi^{\star}[\boldsymbol{\varepsilon}] + \mathcal{O}(\eta^3)$, in the form

$$\Phi_+[\boldsymbol{\varepsilon}] = \int_{\mathbb{R}^2} \frac{1}{2} \tilde{\boldsymbol{\varepsilon}} : \boldsymbol{K} : \tilde{\boldsymbol{\varepsilon}}\, \mathrm{d}\boldsymbol{x} \quad \text{where } \tilde{\boldsymbol{\varepsilon}} = \left\{ \boldsymbol{\varepsilon}(\boldsymbol{x}) + \eta^2 \boldsymbol{K}^{-1} : \boldsymbol{C} :: \nabla^2 \boldsymbol{\varepsilon}(\boldsymbol{x}) \right\}. \tag{3.20}$$

Here, $\boldsymbol{K}^{-1} = \frac{\boldsymbol{I}_2 \otimes \boldsymbol{I}_2}{4(\lambda + \mu)} + \frac{\mathscr{D}}{2\mu}$ is the inverse of the strictly positive quadratic form $\boldsymbol{K}$, satisfying $\boldsymbol{K}^{-1} : \boldsymbol{K} : \boldsymbol{\varepsilon} = \boldsymbol{\varepsilon}$. The proposed energy $\Phi_+$ matches that in (3.12) up to a negligible term, of order $\eta^4$. It also closely resembles that of the laminate in (2.34), except for the fact that the triangular lattice has $\boldsymbol{B} = \boldsymbol{0}$. The positivity of leading-order stiffness $\boldsymbol{K}$ in (3.16) warrants that $\Phi_+[\boldsymbol{\varepsilon}]$ is positive.

The generalized strain $\tilde{\boldsymbol{\varepsilon}} = \{\boldsymbol{\varepsilon} + \eta^2 \boldsymbol{K}^{-1} : \boldsymbol{C} :: \nabla^2 \boldsymbol{\varepsilon}\}$ is obtained in the companion notebook in explicit form as

$$\tilde{\boldsymbol{\varepsilon}} = \left\{ \boldsymbol{\varepsilon} + \frac{\eta^2}{48} \left( \Delta \boldsymbol{\varepsilon} + (\nabla^2 \mathrm{tr}\, \boldsymbol{\varepsilon})^D + \left[ \frac{\mathscr{J} : \nabla^2 \boldsymbol{\varepsilon} : \mathscr{J}^{T[3,1,2]}}{2} \right]^{\mathrm{sym}} \right) \right\}, \tag{3.21}$$

where $(\nabla^2 \mathrm{tr}\, \boldsymbol{\varepsilon})^D = \mathscr{D} :: (\nabla^2 \mathrm{tr}\, \boldsymbol{\varepsilon}) = \nabla^2 \mathrm{tr}\, \boldsymbol{\varepsilon} - \boldsymbol{I}_2 \Delta \mathrm{tr}\, \boldsymbol{\varepsilon}/2$ denotes the deviatoric part of the second gradient of $(\mathrm{tr}\, \boldsymbol{\varepsilon})$. This, together with the expression of $\boldsymbol{K}$ in (3.14), makes the expression (3.20) of $\Phi_+$ fully explicit.

**Note 3.3.** In its first version, the library shoal was implementing the naïve truncation method from Section 3.4, which broke positivity. Along with this paper, we are releasing an update to the library that outputs the homogenized properties $\boldsymbol{K}$, $\boldsymbol{B}$ and $\boldsymbol{C}$ without integrating by parts. This update to the library was used to produce all the homogenization results for the lattices analyzed in this paper (triangular, honeycomb and pantograph).

**Remark 3.4. (geometric compatibility)** A symmetric tensor field $\tilde{\boldsymbol{\varepsilon}}(\boldsymbol{x})$ is said to be *geometrically compatible* if there exists a vector field $\tilde{\boldsymbol{u}}(\boldsymbol{x})$ such that $\tilde{\boldsymbol{\varepsilon}}(\boldsymbol{x}) = (\nabla \tilde{\boldsymbol{u}}(\boldsymbol{x}) + \nabla \tilde{\boldsymbol{u}}^T(\boldsymbol{x}))/2$. In dimension 2, a necessary condition for geometrically compatibility is $\partial \tilde{\varepsilon}_{11}/\partial x_2^2 + \partial \tilde{\varepsilon}_{22}/\partial x_1^2 - 2\, \partial \tilde{\varepsilon}_{12}/(\partial x_1 \partial x_2) = 0$ for all $\boldsymbol{x}$. In the companion notebook, we test this equality using the generalized strain $\tilde{\boldsymbol{\varepsilon}}$ given in (3.21) and find that it is violated at order $\eta^2$. This proves that there does *not* exist a generalized displacement $\tilde{\boldsymbol{u}}(\boldsymbol{x})$ in terms of which the generalized strain could be defined as $\tilde{\boldsymbol{\varepsilon}}(\boldsymbol{x}) = (\nabla \tilde{\boldsymbol{u}}(\boldsymbol{x}) + \nabla \tilde{\boldsymbol{u}}^T(\boldsymbol{x}))/2$.

# 4 A complex lattice: the honeycomb

In this section, we consider the honeycomb lattice in dimension 2, as sketched in Figure 4.1. It is a complex lattice as it contains two different types of nodes (shown in black and in white in the figure) that are not congruent to one another via the generating translations of the lattice. In complex lattices, one needs to introduce microscopic degrees of freedom, denoted as $\boldsymbol{\xi}$ and $\psi_\pm$ here, that capture the relative displacements and rotations of the two types of nodes. Homogenization links the microscopic degrees of freedom to the macroscopic strain $\boldsymbol{\varepsilon}$ and its gradients.

Second-order homogenization of the honeycomb lattice has been carried out in [YAL24], and we follow closely this presentation in Sections 4.1 to 4.4. In Section 4.5, we apply the novel truncation scheme preserving positivity to the non-positive energy obtained in this previous work.

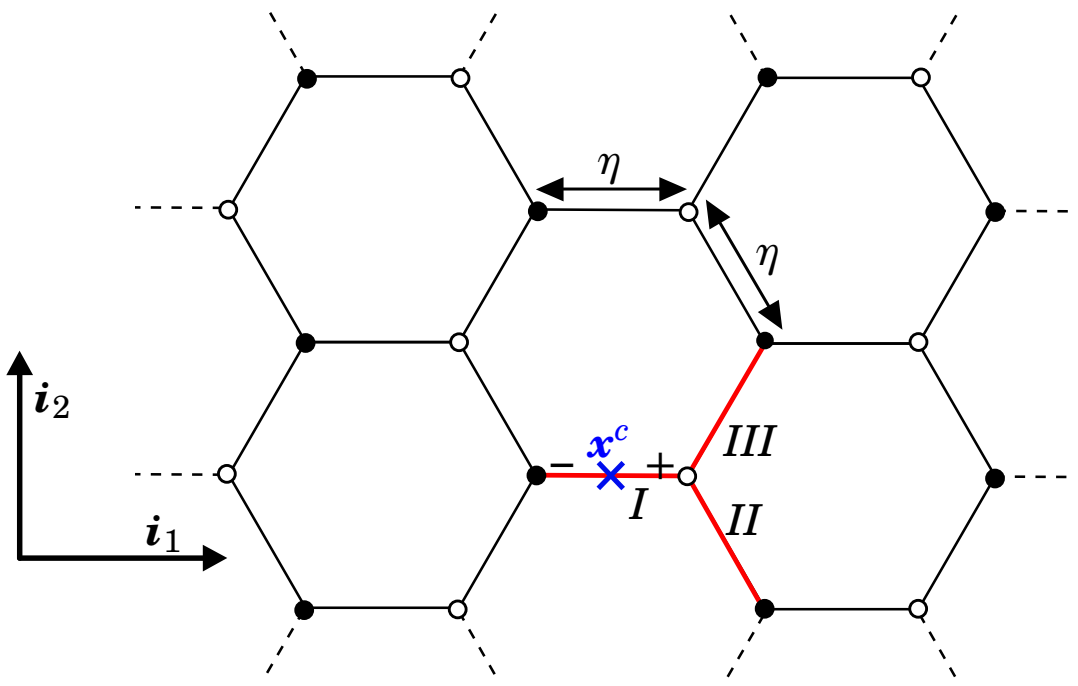

**Figure 4.1.** A 2D infinite periodic honeycomb lattice, composed of identical beams. The red cell, indexed $i$, is the minimal unit-cell and contains three beams of type $\varphi \in \{I, II, III\}$. The center of a particular beam that is analyzed in the text is denoted as $\boldsymbol{x}^c$.

## 4.1 Geometry and elastic energy

We consider an infinite honeycomb lattice, obtained by repeating the unit-cell (show in red in the figure) periodically along the generating vectors $\boldsymbol{i}'_1 = \left(\sqrt{3}\,\boldsymbol{i}_1 + \boldsymbol{i}_2\right)/2$ and $\boldsymbol{i}_2$. The unit cell contains three families of beams of length $\eta$, each family being labelled $\varphi \in \{I, II, III\}$. The beams are considered inextensible.

When this complex lattice is homogenized, the Cauchy–Born rule states that the displacement $\boldsymbol{v}_\beta$ and rotation $\theta_\beta$ of a node having undeformed position $\boldsymbol{x}_\beta$ take different forms depending on the node family, see also Equation [3.4] in [YAL24],

| | displacement $\boldsymbol{v}_\beta$ | rotation $\theta_\beta$ | |
|---|---|---|---|
| white nodes | $\boldsymbol{u}(\boldsymbol{x}_\beta) - \eta\,\boldsymbol{\xi}(\boldsymbol{x}_\beta)$ | $\gamma(\boldsymbol{x}_\beta) + \psi_-(\boldsymbol{x}_\beta)$ | (4.1) |
| black nodes | $\boldsymbol{u}(\boldsymbol{x}_\beta) + \eta\,\boldsymbol{\xi}(\boldsymbol{x}_\beta)$ | $\gamma(\boldsymbol{x}_\beta) + \psi_+(\boldsymbol{x}_\beta)$. | |

Here $\boldsymbol{u}(\boldsymbol{x})$, $\boldsymbol{\xi}(\boldsymbol{x})$, $\gamma(\boldsymbol{x})$ and $\psi_\pm(\boldsymbol{x})$ are smooth functions of the macroscopic variable $\boldsymbol{x}$ that can be expressed as regular series in $\eta$, as earlier in (2.10), $\boldsymbol{u}(\boldsymbol{x})$ being the macroscopic displacement, $\gamma(\boldsymbol{x}) = (\partial u_2/\partial x_1 - \partial u_1/\partial x_2)/2$ the associated rotation, $\boldsymbol{\xi}(\boldsymbol{x})$ the microscopic shift and $\psi_\pm(\boldsymbol{x})$ the microscopic rotation of the black nodes ($\psi_-$) and white notes ($\psi_+$). Even though these functions are smooth, the displacement $\boldsymbol{v}_\beta$ and rotation $\theta_\beta$ in (4.1) vary quickly at the scale of a unit cell due to the alternating signs $\pm$ in front of the microscopic displacement ($\pm\eta\,\boldsymbol{\xi}$ term) and in the microrotation ($\psi_\pm$ terms).

The main task of homogenization is to solve for the microscopic displacement and rotation in terms of the macroscopic strain $\boldsymbol{\varepsilon}(\boldsymbol{x}) = (\nabla\boldsymbol{u} + \nabla\boldsymbol{u}^T)/2$, see (3.3). A perturbative solution is available from Equation [5.10] in [YAL24],

$$
\begin{aligned}
\boldsymbol{\xi}(\boldsymbol{x}) &= -\frac{1}{4}\mathcal{J} : \boldsymbol{\varepsilon}(\boldsymbol{x}) + \boldsymbol{0} \times \eta + \dots \\
\psi_\pm(\boldsymbol{x}) &= 0 \pm \frac{3\,\eta}{4}\operatorname{curl}\left(\mathcal{J} : \boldsymbol{\varepsilon}(\boldsymbol{x})\right) + \dots
\end{aligned}
\tag{4.2}
$$

where $\mathcal{J}$ is the rank-3 tensor characteristic of $D_6$ symmetry introduced in (3.18) and $\operatorname{curl}\boldsymbol{a}(\boldsymbol{x})$ denotes the (scalar) curl of a vector field in dimension 2.

The inextensibility of the beams requires the macroscopic deformation to be incompressible,

$$
\operatorname{tr}\boldsymbol{\varepsilon} = 0, \tag{4.3}
$$

as stated for instance in Equation [5.8] from [YAL24].

The energy of the beam labelled '$I$' in Figure 4.1 is given by linear beam theory as $\frac{1}{2}\frac{EI}{\eta}(\kappa_I^2 + 12\,\tau_I^2)$ where $\eta$ is the beam length, $EI$ is the bending modulus, and $\kappa_I = \theta_+ - \theta_-$ and $\tau_I = \frac{\theta_- + \theta_+}{2} - \frac{(\boldsymbol{v}_+ - \boldsymbol{v}_-)\cdot\boldsymbol{i}_2}{\eta}$ are two strain measures relevant to the linear theory of inextensible beams, see Equation [2.12] in [YAL24]. The elastic energy for the entire lattice can then be written as

$$
\Phi = \sum_{\varphi \in \{I, II, III\}} \sum_{i \in \varphi} \frac{1}{2}\frac{EI}{\eta}(\kappa_\varphi^2 + 12\,\tau_\varphi^2), \tag{4.4}
$$

where the first sum runs over beam families $\varphi$, and the second sum over beams belonging to a particular family, see Equations [2.16] and [2.23] in [YAL24]. The dependence of the strain measures $\kappa_\varphi$ and $\tau_\varphi$ on the particular member $i$ of the family $\varphi$ is implicit in the sum above.

## 4.2 Series solution for the microscopic strain

The strain measure $\kappa_I = \theta_+ - \theta_-$ quantifies the mean curvature imposed by the rotation $\theta_\pm$ of the endpoints. By a similar calculation as earlier in (3.2), but paying attention to the fact that the (–) endpoint is now black whereas the (+) endpoint is white, one can express the strain $\kappa_I$ in the horizontal beam labelled '$I$' in Figure 4.1, having midpoint $\boldsymbol{x}^{\mathrm{c}}$, as

$$\begin{aligned}\kappa_I &= \theta_+ - \theta_- \\ &= \left(\gamma\left(\boldsymbol{x}^{\mathrm{c}} + \frac{\eta}{2}\boldsymbol{i}_1\right) + \psi_-\left(\boldsymbol{x}^{\mathrm{c}} + \frac{\eta}{2}\boldsymbol{i}_1\right)\right) - \left(\gamma\left(\boldsymbol{x}^{\mathrm{c}} - \frac{\eta}{2}\boldsymbol{i}_1\right) + \psi_+\left(\boldsymbol{x}^{\mathrm{c}} - \frac{\eta}{2}\boldsymbol{i}_1\right)\right) \\ &= \eta\,\nabla\gamma(\boldsymbol{x}^{\mathrm{c}}) + \cdots + \psi_-(\boldsymbol{x}^{\mathrm{c}}) + \frac{\eta}{2}\nabla\psi_-(\boldsymbol{x}^{\mathrm{c}})\cdot\boldsymbol{i}_1 + \cdots - \psi_+(\boldsymbol{x}^{\mathrm{c}}) + \frac{\eta}{2}\nabla\psi_+(\boldsymbol{x}^{\mathrm{c}})\cdot\boldsymbol{i}_1 + \cdots\end{aligned} \tag{4.5}$$

The gradient of the macroscopic rotation $\nabla\gamma$ can be worked out as $\nabla\gamma = (\partial\varepsilon_{12}/\partial x_1 - \partial\varepsilon_{11}/\partial x_2)\,\boldsymbol{i}_1 + (-\partial\varepsilon_{12}/\partial x_2 + \partial\varepsilon_{22}/\partial x_1)\,\boldsymbol{i}_2 = \mathscr{G} \therefore \nabla\boldsymbol{\varepsilon}$ in terms of the strain gradient $\nabla\boldsymbol{\varepsilon}$, where $\mathscr{G} = \boldsymbol{i}_1 \otimes \frac{\boldsymbol{i}_1\otimes\boldsymbol{i}_2 + \boldsymbol{i}_2\otimes\boldsymbol{i}_1}{2} \otimes \boldsymbol{i}_1 - \boldsymbol{i}_1\otimes\boldsymbol{i}_1\otimes\boldsymbol{i}_1\otimes\boldsymbol{i}_2 - \boldsymbol{i}_2 \otimes \frac{\boldsymbol{i}_1\otimes\boldsymbol{i}_2 + \boldsymbol{i}_2\otimes\boldsymbol{i}_1}{2} \otimes \boldsymbol{i}_2 + \boldsymbol{i}_2\otimes\boldsymbol{i}_2\otimes\boldsymbol{i}_2\otimes\boldsymbol{i}_1$ is a constant rank-4 tensor, see Appendix A.3 of [YAL24]. On using the solution $(4.2)_2$ for the microscopic shift and rewriting curl $(\mathscr{J} : \boldsymbol{\varepsilon}) = \mathscr{G}' \therefore \nabla\boldsymbol{\varepsilon}$ with the help of another constant rank-4 tensor $\mathscr{G}'$ (whose expression is omitted for the sake of brevity), one obtains the Taylor expansion of $\kappa_I$ as

$$\kappa_I = \left[\boldsymbol{0}\right] : \boldsymbol{\varepsilon}(\boldsymbol{x}^{\mathrm{c}}) + \eta\left[\mathscr{G} - \frac{3}{2}\mathscr{G}'\right] \therefore \nabla\boldsymbol{\varepsilon}(\boldsymbol{x}^{\mathrm{c}}) + \eta^2\left[\boldsymbol{0}\right] :: \nabla^2\boldsymbol{\varepsilon}(\boldsymbol{x}^{\mathrm{c}}) + O(\eta^3). \tag{4.6}$$

Next, we introduce the full set of deformation measures relevant to all 3 families of beams in the form the microscopic strain vector,

$$\boldsymbol{E} = (\ \kappa_I\ \ \tau_I\ \ \kappa_{II}\ \ \tau_{II}\ \ \kappa_{III}\ \ \tau_{III}\ ) \in \mathbb{R}^6. \tag{4.7}$$

A similar calculation as that done above in (4.6) yields the Taylor expansions of the strain measures $\kappa_\varphi$ and $\tau_\varphi$ for each type $\varphi$ of beams. Collecting them, one obtains a Taylor expansion for $\boldsymbol{E} = \boldsymbol{E}(\boldsymbol{x})$ in the same form as earlier in (3.6), where $\boldsymbol{E}(\boldsymbol{x})$ is the smooth function

$$\boldsymbol{E}(\boldsymbol{x}) = \boldsymbol{F}_0 : \boldsymbol{\varepsilon}(\boldsymbol{x}) + \eta\,\boldsymbol{F}_1 \therefore \nabla\boldsymbol{\varepsilon}(\boldsymbol{x}) + \eta^2\boldsymbol{F}_2 :: \nabla^2\boldsymbol{\varepsilon}(\boldsymbol{x}) + \mathcal{O}(\eta^3). \tag{4.8}$$

The tensors $\boldsymbol{F}_0$, $\boldsymbol{F}_1$ and $\boldsymbol{F}_2$ are calculated in symbolic form using the homogenization library shoal in the companion notebook homogenize-inextensible-honeycomb.nb, provided as supplementary material [TAL24b]. The tensors $\boldsymbol{F}_i$ are constant that have no dependence on any parameter.

## 4.3 Series expansion of the energy

By a similar argument as earlier in Equation (3.8), the strain energy $\Phi$ in (4.4) can be continualized as

$$\Phi[\boldsymbol{E}] = \frac{1}{2}\int_{\mathbb{R}^2} \boldsymbol{E}(\boldsymbol{x}) \cdot \boldsymbol{k} \cdot \boldsymbol{E}(\boldsymbol{x})\,\mathrm{d}\boldsymbol{x} \tag{4.9}$$

where $\boldsymbol{k}$ is the diagonal matrix

$$\boldsymbol{k} = \frac{\rho EI}{\eta}\begin{pmatrix} \begin{pmatrix} 1 & \\ & 12 \end{pmatrix} & & \\ & \begin{pmatrix} 1 & \\ & 12 \end{pmatrix} & \\ & & \begin{pmatrix} 1 & \\ & 12 \end{pmatrix} \end{pmatrix}, \tag{4.10}$$

with missing components being zero. The 3 diagonal blocks are each associated with a family of spring, to reflect the ordering conventions in (4.7). The coefficients 1 and 12 making up each block come from Equation (4.4) and $\rho = 2/\left(3\sqrt{3}\,\eta^2\right)$ is the area density of each particular type of beam. The continualized form of the energy (4.9) appears in Equations [3.38,3.39,5.1] in the original paper [YAL24].

For the lattice energy to remain finite in the limit $\eta \to 0$, we require that the bending moduli $EI$ scale as $\eta^3$, *i.e.*, we rewrite $EI = \overline{EI}\,\eta^3$ where the scaled bending modulus $\overline{EI}$ is finite. The coefficient in the right-hand side of (4.10) can then be rewritten as $\rho EI/\eta = 2\,\overline{EI}/\left(3\sqrt{3}\right)$.

By inserting the approximated strain $\boldsymbol{E}(\boldsymbol{x})$ (4.8) into the continuous energy $\Phi[\boldsymbol{E}]$ (4.9), we identify an effective energy for the honeycomb lattice in the form

$$\Phi^\star[\boldsymbol{\varepsilon}^D] = 2\mu\int_{\mathbb{R}^2} \frac{1}{2}\boldsymbol{\varepsilon}^D(\boldsymbol{x}) : \boldsymbol{\varepsilon}^D(\boldsymbol{x}) + \eta^2\left(\frac{1}{2}\nabla\boldsymbol{\varepsilon}^D(\boldsymbol{x}) \therefore \boldsymbol{B} \therefore \nabla\boldsymbol{\varepsilon}^D(\boldsymbol{x}) + \boldsymbol{\varepsilon}^D(\boldsymbol{x}) : \boldsymbol{C} :: \nabla^2\boldsymbol{\varepsilon}^D(\boldsymbol{x})\right)\mathrm{d}\boldsymbol{x} + \mathcal{O}(\eta^3) \tag{4.11}$$

where $\mu$ is the homogenized shear modulus predicted by classical homogenization

$$\mu = 4\sqrt{3}\,\overline{EI}, \tag{4.12}$$

which is obviously positive, $\mu > 0$. By the incompressibility condition (4.3), the macroscopic strain is purely deviatoric, hence the notation $\boldsymbol{\varepsilon} = \boldsymbol{\varepsilon}^D$ in the homogenized energy. There is no term coupling $\boldsymbol{\varepsilon}^D$ and $\nabla\boldsymbol{\varepsilon}^D$ in the expression (4.11) of $\Phi^\star$, as a consequence of the lattice being centro-symmetric.

The tensors $\boldsymbol{B}$ and $\boldsymbol{C}$ are calculated by the library shoal in terms of the $\boldsymbol{F}_i$'s using formulas similar to (2.24) and (3.13). They are given in symbolic form in the companion notebook.

With the help of symbolic calculations, we have further reduced the operators $\boldsymbol{B}$ and $\boldsymbol{C}$ in the companion notebook: their action on the first and second gradient of strain has been rewritten in terms of invariants as

$$\begin{aligned}\nabla\boldsymbol{\varepsilon}^D(\boldsymbol{x}) \therefore \boldsymbol{B} \therefore \nabla\boldsymbol{\varepsilon}^D(\boldsymbol{x}) &= \frac{3}{16}\mathrm{curl}^2(\mathcal{J}:\boldsymbol{\varepsilon}^D) + \frac{1}{48}\|\mathcal{D}:\nabla(\mathcal{J}:\boldsymbol{\varepsilon}^D)\|^2 \\ \boldsymbol{C}::\nabla^2\boldsymbol{\varepsilon}^D(\boldsymbol{x}) &= \frac{1}{16}\left(3\,\Delta\boldsymbol{\varepsilon}^D - 7\,\mathcal{D}:\left[\frac{\mathcal{J}:\nabla^2\boldsymbol{\varepsilon}^D:\mathcal{J}^{T[3,1,2]}}{2}\right]\right),\end{aligned} \tag{4.13}$$

where $\mathcal{J}$ is given in (3.18), $\mathcal{D}$ and $T[3,1,2]$ are the deviatoric projector and generalized transpose already appearing in (3.17), respectively, $\Delta$ is the Laplacian and the square norm $\|\boldsymbol{c}\|^2$ of a (deviatoric) tensor $\boldsymbol{c}$ is defined as $\|\boldsymbol{c}\|^2 = \boldsymbol{c}:\boldsymbol{c}$.

The positivity of the quadratic form $\boldsymbol{B}$ is evident from $(4.13)_1$,

$$\nabla\boldsymbol{\varepsilon}^D(\boldsymbol{x}) \therefore \boldsymbol{B} \therefore \nabla\boldsymbol{\varepsilon}^D(\boldsymbol{x}) \geqslant 0. \tag{4.14}$$

## 4.4 Naïve truncation of the energy

The original analysis of honeycomb lattices [YAL24] was based on an early version of the library shoal which calculated the expression (4.11) internally but carried out an integration by parts and the naïve truncation without exposing the form $\Phi^\star$ nor the tensors $\boldsymbol{B}$ and $\boldsymbol{C}$ to the user: the homogenized energy reported in Equation [5.9] of this paper [YAL24] took the form (analogous to (3.19))

$$\Phi^\ddagger[\boldsymbol{\varepsilon}^D] = 2\,\mu\int_{-\infty}^{+\infty}\left(\frac{1}{2}\boldsymbol{\varepsilon}^D:\boldsymbol{\varepsilon}^D + \frac{\eta^2}{2\times 8}\left(-7\,\nabla\boldsymbol{\varepsilon}^D \therefore \nabla\boldsymbol{\varepsilon}^D + \frac{22}{3}\|\nabla\boldsymbol{\varepsilon}^D:\boldsymbol{I}_2\|^2 + 2\,\mathrm{div}^2(\mathcal{J}:\boldsymbol{\varepsilon}^D)\right)\right)\mathrm{d}\boldsymbol{x}. \tag{4.15}$$

The strain-gradient contribution is non-positive due to the negative coefficient $(-7)$, as noted by the authors in their Remark 17.

## 4.5 Truncation preserving positivity

Applying instead the positivity-preserving truncation scheme from Section 2.6 to (4.11), we obtain an alternative expression of the homogenized energy that is consistent to order $\eta^2$ included,

$$\Phi_+[\boldsymbol{\varepsilon}^D] = 2\,\mu\int_{\mathbb{R}^2}\left(\begin{array}{l}\frac{1}{2}\left\{\boldsymbol{\varepsilon}^D(\boldsymbol{x}) + \eta^2\boldsymbol{C}::\nabla^2\boldsymbol{\varepsilon}^D(\boldsymbol{x})\right\}:\left\{\boldsymbol{\varepsilon}^D(\boldsymbol{x}) + \eta^2\boldsymbol{C}::\nabla^2\boldsymbol{\varepsilon}^D(\boldsymbol{x})\right\} \\ \quad + \frac{\eta^2}{2}\nabla\boldsymbol{\varepsilon}^D(\boldsymbol{x}) \therefore \boldsymbol{B} \therefore \nabla\boldsymbol{\varepsilon}^D(\boldsymbol{x})\end{array}\right)\mathrm{d}\boldsymbol{x}. \tag{4.16}$$

The quantity $\{\ldots\}$ inside the curly braces is a generalized strain measure. Equation $(4.13)_2$ shows that the strain correction $\eta^2\,\boldsymbol{C}::\nabla^2\boldsymbol{\varepsilon}^D(\boldsymbol{x})$ is a *deviatoric* rank-2 tensor (*i.e.*, it is symmetric and traceless) like the leading-order strain $\boldsymbol{\varepsilon}^D$.

A fully explicit expression of $\Phi_+$ is obtained by combining the expression above with (4.13), which yields

$$\Phi_+[\boldsymbol{\varepsilon}^D] = 2\,\mu\int_{\mathbb{R}^2}\left(\begin{array}{l}\frac{1}{2}\left\|\boldsymbol{\varepsilon}^D + \frac{\eta^2}{16}\left(3\,\Delta\boldsymbol{\varepsilon}^D - 7\,\mathcal{D}:\left[\frac{\mathcal{J}:\nabla^2\boldsymbol{\varepsilon}^D:\mathcal{J}^{T[3,1,2]}}{2}\right]\right)\right\|^2 \\ \quad + \frac{\eta^2}{2}\left(\frac{3}{16}\mathrm{curl}^2(\mathcal{J}:\boldsymbol{\varepsilon}^D) + \frac{1}{48}\|\mathcal{D}:\nabla(\mathcal{J}:\boldsymbol{\varepsilon}^D)\|^2\right)\end{array}\right)\mathrm{d}\boldsymbol{x}, \tag{4.17}$$

This functional is clearly positive.

**Remark 4.1. (geometric compatibility)** By the same method as for the triangular lattice, see Remark 3.4 at the end of section 3, it has been checked in the companion notebook that the generalized strain $\left\{\boldsymbol{\varepsilon}^D(\boldsymbol{x}) + \eta^2\boldsymbol{C}::\nabla^2\boldsymbol{\varepsilon}^D(\boldsymbol{x})\right\}$ is *not* geometrically compatible.

# 5 Connecting with Cholesky decomposition

In this section, we connect with the LDLT Cholesky decomposition, a classical method used in numerical analysis [Str07]. The Cholesky decomposition takes a symmetric positive matrix $\boldsymbol{t}$ on input and decomposes it *iteratively* into $\boldsymbol{t} = \boldsymbol{L}\cdot\boldsymbol{D}\cdot\boldsymbol{L}^T$, where $\boldsymbol{L}$ is a lower-triangular matrix filled with 1's on the diagonal, and $\boldsymbol{D}$ is a diagonal matrix containing only positive entries, $\boldsymbol{D} \geqslant 0$.

We introduce the following variant of the Cholesky decomposition:

- We start from a symbolic matrix $\boldsymbol{t}$ whereas the original Cholesky method works on numerical matrices.
- We carry out only the first steps of the Cholesky decomposition: when homogenizing to order $\eta^p$ included ($p = 2$ in our examples), we stop after $p + 1$ iterations.
- In our variant, the input matrix $\boldsymbol{t}$ is defined by *blocks* and the Cholesky decomposition is carried out at the block level and not at the level of the individual entries of $\boldsymbol{t}$. This enables us to handle the common situation where there are multiple scalar measures of macroscopic strain—the laminate example from Section 2, for instance, uses a strain $\boldsymbol{E}$ having 2 components, see (2.4).

The Cholesky decomposition helps extending the truncation method proposed so far in two important ways:

- to non-centro-symmetric structures (the examples analyzed so far were all centro-symmetric, a particular case in which our truncation method simply requires moving the cross terms of the form $C\,\varepsilon\,\varepsilon''$ into the leading order $K\,\varepsilon^2/2$, see for instance Section 2.6).
- to arbitrary homogenization order (even though emphasis is on second-order homogenization here).

## 5.1 Cholesky decomposition by blocks

The input to our variant of the Cholesky decomposition is symmetric matrix $\boldsymbol{t}$ written in block-matrix notation as

$$\boldsymbol{t} = \begin{pmatrix} \boldsymbol{K} & \underline{\boldsymbol{A}} & \underline{\underline{\boldsymbol{C}}} & \cdots \\ \underline{\boldsymbol{A}}^T & \underline{\underline{\boldsymbol{B}}} & \cdots & \\ \underline{\underline{\boldsymbol{C}}}^T & \cdots & & \\ \cdots & & & \end{pmatrix}, \tag{5.1}$$

where the ellipses represents blocks that we do not need to resolve as they represent higher-order effects. Obviously, the diagonal blocks $\boldsymbol{K}$, $\underline{\underline{\boldsymbol{B}}}$, etc. have to be symmetric.

Anticipating on what comes next, we annotate the various blocks with multiple underlines to denote the iteration at which the various blocks enter into the procedure: $\boldsymbol{K}$ will be used in the first iteration (numbered 0), $\underline{\boldsymbol{A}}$ in the second iteration (numbered 1) and $\underline{\underline{\boldsymbol{B}}}$ and $\underline{\underline{\boldsymbol{C}}}$ in the third iteration (numbered 2). A key remark is that the iteration numbers match the orders of the energy terms associated with the different blocks when $\boldsymbol{t}$ is interpreted as the matrix of elastic coefficients—see Equations (2.19) and (2.22) for example. Indeed, the $\boldsymbol{K}$ block gives rise to an energy contribution $\boldsymbol{\varepsilon}\cdot\boldsymbol{K}\cdot\boldsymbol{\varepsilon}/2$ of order $\eta^0$, the $\underline{\boldsymbol{A}}$ block gives rise to an energy contribution $\boldsymbol{\varepsilon}\cdot\boldsymbol{A}\cdot(\eta\,\boldsymbol{\varepsilon}')$ of order $\eta^1$, the $\underline{\underline{\boldsymbol{B}}}$ and $\underline{\underline{\boldsymbol{C}}}$ blocks give rise to energy contributions $(\eta\,\boldsymbol{\varepsilon}')\cdot\underline{\underline{\boldsymbol{B}}}\cdot(\eta\,\boldsymbol{\varepsilon}')/2+\boldsymbol{\varepsilon}\cdot\underline{\underline{\boldsymbol{C}}}\cdot(\eta^2\,\boldsymbol{\varepsilon}'')$ of order $\eta^2$, etc. This match in the orders of magnitude of the blocks in the matrix $\boldsymbol{t}$ produced by homogenization, and in the iteration number of the Cholesky decomposition makes the latter perfectly suited to the former.

We assume that the top-left (symmetric) $\boldsymbol{K}$ block is positive definite, and that the matrix $\boldsymbol{t}$ is positive but not necessarily definite. The only positivity properties which we will use in second-order homogenization are

$$\boldsymbol{K}>0, \qquad \begin{pmatrix} \boldsymbol{K} & \underline{\boldsymbol{A}} \\ \underline{\boldsymbol{A}}^T & \underline{\underline{\boldsymbol{B}}} \end{pmatrix} \geqslant 0. \tag{5.2}$$

For homogenization at order $p$, we would assume that the first $p$ square super-blocks starting from the top-left corner of $\boldsymbol{t}$ are all positive.

Following Cholesky, we seek a decomposition of $\boldsymbol{t}$ in the form

$$\boldsymbol{t} = \begin{pmatrix} \boldsymbol{I} & \boldsymbol{0} & \boldsymbol{0} & \cdots \\ \underline{\boldsymbol{L}}_1 & \boldsymbol{I} & \boldsymbol{0} & \cdots \\ \underline{\underline{\boldsymbol{L}}}_2 & \cdots & & \\ \cdots & & & \end{pmatrix} \cdot \begin{pmatrix} \boldsymbol{D}_0 & \boldsymbol{0} & \boldsymbol{0} & \cdots \\ \boldsymbol{0} & \underline{\underline{\boldsymbol{D}}}_1 & \cdots & \\ \boldsymbol{0} & \cdots & & \\ \cdots & & & \end{pmatrix} \cdot \begin{pmatrix} \boldsymbol{I} & \boldsymbol{0} & \boldsymbol{0} & \cdots \\ \underline{\boldsymbol{L}}_1 & \boldsymbol{I} & \boldsymbol{0} & \cdots \\ \underline{\underline{\boldsymbol{L}}}_2 & \cdots & & \\ \cdots & & & \end{pmatrix}^T, \tag{5.3}$$

where the first matrix is lower-triangular by blocks and has only identity blocks on the diagonal, the third matrix is the transpose of the first, and the second matrix is block-diagonal.

Calculating the matrix products on the right-hand side of (5.3) and identifying with (5.1), one obtains the following relations: at iteration 0, $\boldsymbol{D}_0=\boldsymbol{K}$; at iteration 1, $\boldsymbol{D}_0\cdot\underline{\boldsymbol{L}}_1^T=\underline{\boldsymbol{A}}$; at iteration 2, $\boldsymbol{D}_0\cdot\underline{\underline{\boldsymbol{L}}}_2^T=\underline{\underline{\boldsymbol{C}}}$ and $\underline{\boldsymbol{L}}_1\cdot\boldsymbol{D}_0\cdot\underline{\boldsymbol{L}}_1^T+\underline{\underline{\boldsymbol{D}}}_1=\underline{\underline{\boldsymbol{B}}}$, etc. The pieces of the LDLT decomposition can be solved from these relations uniquely and the solution is shown in Table 5.1. Note that we have correctly anticipated the iteration number, shown by the number of underlines, at which the quantities $\boldsymbol{L}_i$ and $\boldsymbol{D}_i$ are solved (left-hand sides in the table) and at which the blocks of $\boldsymbol{t}$ are first used (in the right-hand sides in the table).

| iteration | 0 | 1 | 2 |
|---|---|---|---|
| moduli | $\boldsymbol{D}_0=\boldsymbol{K}$ | | $\underline{\underline{\boldsymbol{D}}}_1=\underline{\underline{\boldsymbol{B}}}-\underline{\boldsymbol{A}}^T\cdot\boldsymbol{K}^{-T}\cdot\underline{\boldsymbol{A}}$ |
| coefficients | | $\underline{\boldsymbol{L}}_1=(\boldsymbol{K}^{-1}\cdot\underline{\boldsymbol{A}})^T$ | $\underline{\underline{\boldsymbol{L}}}_2=(\boldsymbol{K}^{-1}\cdot\underline{\underline{\boldsymbol{C}}})^T$ |

**Table 5.1.** Solution for the Cholesky decomposition.

We now apply the quadratic form associated to $\boldsymbol{t}$ in (5.1) to a vector written in block-vector notation as $\begin{pmatrix} \boldsymbol{\varepsilon}^{(0)} \\ \eta\,\boldsymbol{\varepsilon}^{(1)} \\ \cdots \end{pmatrix}$. Each of the subvectors $\boldsymbol{\varepsilon}^{(k)}$ is a flattened (rank-1) version of the rank-$(k+2)$ tensor $\nabla^k\boldsymbol{\varepsilon}$ representing the $k$-th gradient of the macroscopic strain, as earlier in Equation (2.12). Using the Cholesky decomposition (5.3), we have

$$\frac{1}{2}\begin{pmatrix} \boldsymbol{\varepsilon}^{(0)} \\ \eta\,\boldsymbol{\varepsilon}^{(1)} \\ \eta^2\,\boldsymbol{\varepsilon}^{(2)} \\ \cdots \end{pmatrix} \cdot \boldsymbol{t} \cdot \begin{pmatrix} \boldsymbol{\varepsilon}^{(0)} \\ \eta\,\boldsymbol{\varepsilon}^{(1)} \\ \eta^2\,\boldsymbol{\varepsilon}^{(2)} \\ \cdots \end{pmatrix} = \frac{1}{2}\left(\hat{\boldsymbol{\varepsilon}}^{(0)}\cdot\boldsymbol{D}_0\cdot\hat{\boldsymbol{\varepsilon}}^{(0)}+\eta^2\,\hat{\boldsymbol{\varepsilon}}^{(1)}\cdot\underline{\underline{\boldsymbol{D}}}_1\cdot\hat{\boldsymbol{\varepsilon}}^{(1)}+\cdots\right) \tag{5.4}$$

where the generalized strains $\hat{\boldsymbol{\varepsilon}}^{(0)}$, $\hat{\boldsymbol{\varepsilon}}^{(1)}$, ... are defined as

$$\begin{aligned}\begin{pmatrix}\hat{\boldsymbol{\varepsilon}}^{(0)}\\ \eta\,\hat{\boldsymbol{\varepsilon}}^{(1)}\\ \eta^2\,\hat{\boldsymbol{\varepsilon}}^{(2)}\\ \cdots\end{pmatrix} &= \begin{pmatrix}\boldsymbol{I} & \boldsymbol{0} & \boldsymbol{0} & \cdots\\ \underline{\boldsymbol{L}}_1 & \boldsymbol{I} & \boldsymbol{0} & \cdots\\ \underline{\underline{\boldsymbol{L}}}_2 & \cdots & & \\ \cdots & & & \end{pmatrix}^T \cdot \begin{pmatrix}\boldsymbol{\varepsilon}^{(0)}\\ \eta\,\boldsymbol{\varepsilon}^{(1)}\\ \eta^2\,\boldsymbol{\varepsilon}^{(2)}\\ \cdots\end{pmatrix}\\ &= \begin{pmatrix}\boldsymbol{\varepsilon}^{(0)} & + & \eta\,\underline{\boldsymbol{L}}_1^T\cdot\boldsymbol{\varepsilon}^{(1)} & + & \eta^2\,\underline{\underline{\boldsymbol{L}}}_2^T\cdot\boldsymbol{\varepsilon}^{(2)} & + & \cdots\\ & & \eta\,\boldsymbol{\varepsilon}^{(1)} & + & \cdots & & \\ & & & & \cdots & & \end{pmatrix}.\end{aligned} \tag{5.5}$$

In the right-hand side of (5.4), the matrices $\boldsymbol{D}_i$ are interpreted as homogenized moduli (satisfying the positivity properties established in the forthcoming Section 5.2) and the matrices $\boldsymbol{L}_i$ appearing in (5.5) contain the coefficients entering in the generalized strains.

## 5.2 Positivity of the homogenized moduli

By the same argument as in the original Cholesky decomposition, the positivity of $\boldsymbol{t}$ in (5.2) carries over to the blocks $\boldsymbol{D}_i$:

$$\boldsymbol{D}_0 > 0, \qquad \underline{\underline{\boldsymbol{D}}}_1 \geqslant 0, \qquad \ldots \tag{5.6}$$

This positivity property holds in the sense of quadratic forms, $\boldsymbol{D}_0$ being positive definite while the other $\boldsymbol{D}_i$'s are positive but not necessarily definite.

The proof goes as follows. Inserting (5.1) into (5.3) and extracting the top-left $2\times 2$ super-block (or $p\times p$ super-block at higher order), we get

$$\begin{aligned}\begin{pmatrix}\boldsymbol{K} & \underline{\boldsymbol{A}}\\ \underline{\boldsymbol{A}}^T & \underline{\underline{\boldsymbol{B}}}\end{pmatrix} &= \begin{pmatrix}\boldsymbol{I} & \boldsymbol{0} & \boldsymbol{0} & \cdots\\ \underline{\boldsymbol{L}}_1 & \boldsymbol{I} & \boldsymbol{0} & \cdots\end{pmatrix} \cdot \begin{pmatrix}\boldsymbol{D}_0 & \boldsymbol{0} & \boldsymbol{0} & \cdots\\ \boldsymbol{0} & \underline{\underline{\boldsymbol{D}}}_1 & \cdots & \\ \boldsymbol{0} & \cdots & & \\ \cdots & & & \end{pmatrix} \cdot \begin{pmatrix}\boldsymbol{I} & \boldsymbol{0} & \boldsymbol{0} & \cdots\\ \underline{\boldsymbol{L}}_1 & \boldsymbol{I} & \boldsymbol{0} & \cdots\end{pmatrix}^T\\ &= \begin{pmatrix}\boldsymbol{I} & \boldsymbol{0}\\ \underline{\boldsymbol{L}}_1 & \boldsymbol{I}\end{pmatrix} \cdot \begin{pmatrix}\boldsymbol{D}_0 & \boldsymbol{0}\\ \boldsymbol{0} & \underline{\underline{\boldsymbol{D}}}_1\end{pmatrix} \cdot \begin{pmatrix}\boldsymbol{I} & \boldsymbol{0}\\ \underline{\boldsymbol{L}}_1 & \boldsymbol{I}\end{pmatrix}^T.\end{aligned} \tag{5.7}$$

Multiplying by the inverse of the triangular matrices appearing on both the left and right, we obtain

$$\begin{pmatrix}\boldsymbol{D}_0 & \boldsymbol{0}\\ \boldsymbol{0} & \underline{\underline{\boldsymbol{D}}}_1\end{pmatrix} = \begin{pmatrix}\boldsymbol{I} & \boldsymbol{0}\\ -\underline{\boldsymbol{L}}_1 & \boldsymbol{I}\end{pmatrix} \cdot \begin{pmatrix}\boldsymbol{K} & \underline{\boldsymbol{A}}\\ \underline{\boldsymbol{A}}^T & \underline{\underline{\boldsymbol{B}}}\end{pmatrix} \cdot \begin{pmatrix}\boldsymbol{I} & \boldsymbol{0}\\ -\underline{\boldsymbol{L}}_1 & \boldsymbol{I}\end{pmatrix}^T \tag{5.8}$$

which proves (5.6) using (5.2).

## 5.3 Centro-symmetry as a particular case

In the special case where the unit cell is centro-symmetric, all blocks having an odd number of underlines vanish. In the context of second-order homogenization, the $\underline{\boldsymbol{A}}$ block is the only one vanishing,

$$\underline{\boldsymbol{A}} = \boldsymbol{0}. \tag{5.9}$$

The solution to the Cholesky method in Table 5.1 then boils down to $\boldsymbol{D}_0 = \boldsymbol{K}$, $\boldsymbol{L}_1 = \boldsymbol{0}$, $\boldsymbol{D}_1 = \boldsymbol{B}$ and $\boldsymbol{L}_2 = (\boldsymbol{K}^{-1}\cdot\underline{\underline{\boldsymbol{C}}})^T$, and the decomposition (5.4–5.5) takes the special form

$$\begin{aligned}&\frac{1}{2}\begin{pmatrix}\boldsymbol{\varepsilon}^{(0)}\\ \eta\,\boldsymbol{\varepsilon}^{(1)}\\ \eta^2\,\boldsymbol{\varepsilon}^{(2)}\\ \cdots\end{pmatrix} \cdot \begin{pmatrix}\boldsymbol{K} & \boldsymbol{0} & \underline{\underline{\boldsymbol{C}}} & \cdots\\ \boldsymbol{0} & \underline{\underline{\boldsymbol{B}}} & \cdots & \\ \underline{\underline{\boldsymbol{C}}}^T & \cdots & & \\ \cdots & & & \end{pmatrix} \cdot \begin{pmatrix}\boldsymbol{\varepsilon}^{(0)}\\ \eta\,\boldsymbol{\varepsilon}^{(1)}\\ \eta^2\,\boldsymbol{\varepsilon}^{(2)}\\ \cdots\end{pmatrix}\\ &\qquad = \frac{1}{2}\begin{pmatrix}\{\boldsymbol{\varepsilon}^{(0)} + \eta^2\boldsymbol{K}^{-1}\cdot\underline{\underline{\boldsymbol{C}}}\cdot\boldsymbol{\varepsilon}^{(2)} + \cdots\}\cdot\boldsymbol{K}\cdot\{\boldsymbol{\varepsilon}^{(0)} + \eta^2\boldsymbol{K}^{-1}\cdot\underline{\underline{\boldsymbol{C}}}\cdot\boldsymbol{\varepsilon}^{(2)} + \cdots\}\\ \quad + \{\eta\,\boldsymbol{\varepsilon}^{(1)} + \cdots\}\cdot\underline{\underline{\boldsymbol{B}}}\cdot\{\eta\,\boldsymbol{\varepsilon}^{(1)} + \cdots\}\\ \quad + \cdots\end{pmatrix}\end{aligned} \tag{5.10}$$

This matches the form of the positive energy functionals which we have obtained for the centro-symmetric examples covered so far, namely the laminate in Equation (2.34), the triangular lattice in Equation (3.20), having $\underline{\underline{\boldsymbol{B}}} = \boldsymbol{0}$, and the honeycomb lattice in Equation (4.17). The positivity of the moduli $\boldsymbol{D}_0 = \boldsymbol{K} > 0$ and $\underline{\underline{\boldsymbol{D}}}_1 = \underline{\underline{\boldsymbol{B}}} \geqslant 0$ follows directly from the assumptions in the centro-symmetric case, see (5.2).

In the centro-symmetric case, the Cholesky decomposition therefore reduces to the positivity-preserving truncation method used so far.

# 6 A non-symmetric pantograph

In this section, we analyze a variant of the classical pantograph truss, shown in Figure 6.1. In this variant, the springs shown in black in the figure are stiffer than the springs shown in blue, which breaks the left-right symmetry and makes the lattice non-centro-symmetric. As a result, the homogenized one-dimensional model couples the strain $\varepsilon(x)$ and the strain gradient $\varepsilon'(x)$, as we will see.

The homogenized one-dimensional energy is derived using standard high-order homogenization method first. Next, we apply the Cholesky decomposition from Section 5 to the resulting energy to truncate it in a way that preserves positivity.

Spring constants $2k/\eta$ and $k/\eta$ are assigned to the black and blue springs, respectively. The vertical springs shown by dashed lines in Figure 6.1 are assigned a spring constants $k\chi/\eta$, where $\chi$ is the stiffness contrast parameter. Pantographs are known to feature a floppy mode in the limit $\chi \to 0$ and this floppy mode is coupled to macroscopic stretching [S18b, BGP17, DLSS22, SAI11]. When we homogenize the pantograph, the parameter $\chi$ is treated as a symbol whose value remains unspecified. We discuss the finite-contrast case $\chi = \mathcal{O}(1)$ first, and the high-contrast limit $\chi \to 0$ next.

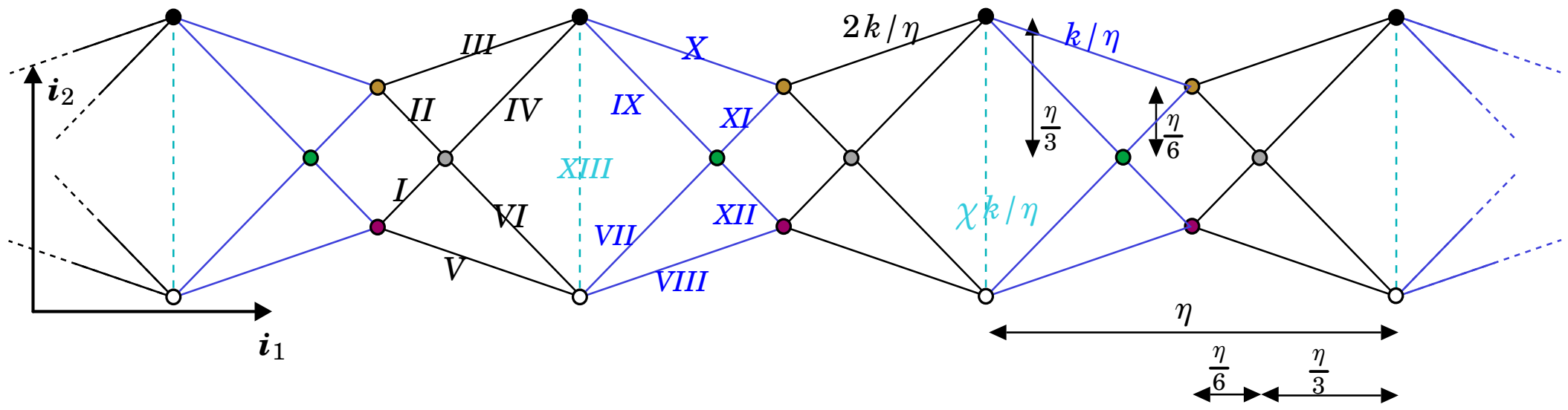

**Figure 6.1.** A non-symmetric pantograph. The black springs have been made stiffer than the blue springs so as to break the mirror-symmetry with respect to a vertical plane. The spring families $\varphi$ are shown using Roman numerals. The spring constant $k\chi\eta^{-1}$ of the vertical springs (dashed line) is parameterized by $\chi$. In the limit of large contrast, $\chi \to 0$, the pantograph displays a soft mode whereby all right triangles move rigidly and the pantograph stretches or contracts horizontally.

## 6.1 Summary of homogenization

The pantograph shown has 13 different types of springs, labelled using Roman numerals, and 6 types of nodes (Bravais sub-lattices), shown using different colors. Except for the choice of spring constants, this pantograph is identical to that studied in section 7.5 of [YAL24]. We homogenize it automatically again using the shoal library in the companion notebook homogenize-pantograph-non-symmetric.nb, provided as supplementary material [TAL24b]. The homogenization follows the same steps as that of the honeycomb (Section 4) and is almost identical to that presented in Section 7.5 of [YAL24] and we limit ourselves to a summary of the main results here.

The homogenization limit $\eta \to 0$ is approached as follows. Working with a symbolic value of the contrast parameter $\chi$, the library calculates the expressions of the microscopic shifts $\boldsymbol{\xi}_\varphi$ of the different Bravais sub-lattices (indexed by $\varphi = I,\dots,XIII$) as an expansion in terms of the successive derivatives of the one-dimensional macroscopic strain $\varepsilon(x) = u'(x)$, where $u(x)$ is the (scalar) macroscopic displacement. Having determined the microscopic shifts, it proceeds to derive the expressions of the rescaled change of length $\varepsilon_\varphi$ of the different types of spring. They are collected in a vector $\boldsymbol{E} = (\,\varepsilon_I \;\cdots\; \varepsilon_{XIII}\,) \in \mathbb{R}^{13}$. Using Taylor expansions, it is written in terms of the macroscopic strain $\varepsilon(x)$ in a similar form as earlier in (4.8),

$$\boldsymbol{E}(x,\chi) = \varepsilon(x)\,\boldsymbol{F}_0(\chi) + \eta\,\varepsilon'(x)\,\boldsymbol{F}_1(\chi) + \eta^2\,\varepsilon''(x)\,\boldsymbol{F}_2(\chi) + \mathcal{O}(\eta^3). \tag{6.1}$$

The strain relocalization vectors $\boldsymbol{F}_0(\chi) = \left(\,\frac{33\sqrt{2}\,\chi}{69+181\,\chi}\;\cdots\,\right)$, $\boldsymbol{F}_1(\chi) = \left(\,-\frac{3\,(9\,\chi+8)}{2\sqrt{2}\,(181\,\chi+69)}\;\cdots\,\right)$ and $\boldsymbol{F}_2(\chi) = \left(\,\frac{60-121\,\chi}{432\sqrt{2}\,(181\,\chi+69)}\;\cdots\,\right)$ are output by the library, see the companion notebook.

The (discrete) strain energy of the lattice can be approximated by the integral

$$\Phi[\boldsymbol{E}] = \frac{1}{2}\int_{-\infty}^{+\infty} \boldsymbol{E}(x,\chi)\cdot\boldsymbol{k}(\chi)\cdot\boldsymbol{E}(x,\chi)\,\mathrm{d}x, \tag{6.2}$$

where

$$\boldsymbol{k}(\chi) = k\,\mathrm{diag}(2,\cdots,2,1,\cdots,1,\chi) \tag{6.3}$$

is the matrix filled with the spring constants $2k/\eta$, $k/\eta$ and $\chi k/\eta$ on the diagonal, times the density $\rho = \eta^{-1}$ of each particular type of spring per unit length, times a factor $\eta^2$ reflecting the convention that the change of the length of springs is $\eta\boldsymbol{E}$ and not just $\boldsymbol{E}$.

Inserting (6.1) into (6.2), one obtains an expansion of the energy of the form

$$\Phi^\star[\varepsilon] = \int_{-\infty}^{+\infty} \frac{1}{2}\begin{pmatrix}\varepsilon(x)\\ \eta\,\varepsilon'(x)\\ \eta^2\,\varepsilon''(x)\end{pmatrix}\cdot\boldsymbol{t}(\chi)\cdot\begin{pmatrix}\varepsilon(x)\\ \eta\,\varepsilon'(x)\\ \eta^2\,\varepsilon''(x)\end{pmatrix}\mathrm{d}x + \mathcal{O}(\eta^3) \tag{6.4}$$

where $\boldsymbol{t}$ is a symmetric $3\times3$ matrix

$$\boldsymbol{t}(\chi) = \begin{pmatrix} K(\chi) & A(\chi) & C(\chi)\\ A(\chi) & B(\chi) & \cdots\\ C(\chi) & \cdots & \cdots \end{pmatrix}, \tag{6.5}$$

whose (non-asymptotic) higher-order contributions, denoted as ellipses, are ignored here. The coefficients of $\boldsymbol{t}$ are given by similar formulas as earlier in (2.24),

| order | diagonal blocks | off-diagonal blocks |
|---|---|---|
| $\eta^0$ | $K(\chi) = \boldsymbol{F}_0(\chi)\cdot\boldsymbol{k}(\chi)\cdot\boldsymbol{F}_0(\chi)$ | |
| $\eta^1$ | | $A(\chi) = \boldsymbol{F}_0(\chi)\cdot\boldsymbol{k}(\chi)\cdot\boldsymbol{F}_1(\chi)$ |
| $\eta^2$ | $B(\chi) = \boldsymbol{F}_1(\chi)\cdot\boldsymbol{k}(\chi)\cdot\boldsymbol{F}_1(\chi)$ | $C(\chi) = \boldsymbol{F}_0(\chi)\cdot\boldsymbol{k}(\chi)\cdot\boldsymbol{F}_2(\chi)$ |

(6.6)

Their expressions and their particular values for $\chi=1$ and $\chi=0$ are listed in Table 6.1.

Note the presence of the $A(\chi)$ term coupling the effective strain $\varepsilon$ and its gradient $\varepsilon'$, which is a consequence of the broken symmetry.

| | $K(\chi)/k$ | $A(\chi)/k$ | $B(\chi)/k$ | $C(\chi)/k$ |
|---|---|---|---|---|
| general case | $\frac{276\chi}{69+181\chi}$ | $\frac{2\chi(69+85\chi)}{(69+181\chi)^2}$ | $\frac{552-19\chi+638\chi^2}{(69+181\chi)^2}$ | $\frac{\chi(1149+17242\chi)}{18(69+181\chi)^2}$ |
| $\chi=1$ | 1.104 | $4.928\times10^{-3}$ | $1.874\times10^{-2}$ | $1.635\times10^{-2}$ |
| $\chi=0$ | 0 | 0 | $1.159\times10^{-1}$ | 0 |

**Table 6.1.** Scaled values of the homogenized elastic constants of the pantograph: general case, case of low elastic contrast ($\chi=1$) and limit of infinite elastic contrast discussed in Section 6.4 ($\chi=0$).

Since $\boldsymbol{k}(\chi)\geqslant 0$ is a positive quadratic form and $\boldsymbol{t}(\chi)$ in (6.5–6.6) can be rewritten as $\boldsymbol{t}(\chi)=(\boldsymbol{F}_0(\chi)\ \boldsymbol{F}_1(\chi)\ \boldsymbol{F}_2(\chi))^T\cdot\boldsymbol{k}(\chi)\cdot(\boldsymbol{F}_0(\chi)\ \boldsymbol{F}_1(\chi)\ \boldsymbol{F}_2(\chi))$, we conclude that $\boldsymbol{t}(\chi)\geqslant 0$ is a positive quadratic form. In particular, its top-left $2\times 2$ sub-block is a positive quadratic form: both the first diagonal entry $K$ and the determinant of the latter are positive,

$$\forall\chi>0,\qquad K(\chi)>0\quad\text{and}\quad K(\chi)\,B(\chi)\geqslant A^2(\chi).\tag{6.7}$$

The *strict* positivity of $K(\chi)=\boldsymbol{F}_0(\chi)\cdot\boldsymbol{k}(\chi)\cdot\boldsymbol{F}_0(\chi)$ comes from the fact that $\boldsymbol{k}(\chi)>0$ and $\boldsymbol{F}_0(\chi)\neq\boldsymbol{0}$ for $\chi>0$. These inequalities can also be checked directly using the expressions in Table 6.1.

In the limit $\chi\to 0$, $K(\chi)\to 0$ and the pantograph becomes soft, as discussed in Section 6.4 below.

## 6.2 Naïve truncation of the energy

When the energy $\Phi^\star[\varepsilon]$ (6.4) is truncated beyond order $\eta^2$, we get

$$\Phi^\dagger[\varepsilon]=\int_{-\infty}^{+\infty}\frac{1}{2}\left(K(\chi)\,\varepsilon^2+2\,\eta A(\chi)\,\varepsilon\,\varepsilon'+\eta^2\left(B(\chi)\,\varepsilon'^2+2\,C(\chi)\,\varepsilon\,\varepsilon''\right)\right)\mathrm{d}x.\tag{6.8}$$

Both the terms $A\,\varepsilon\,\varepsilon'=\frac{\mathrm{d}}{\mathrm{d}x}\left(\frac{A\varepsilon^2}{2}\right)$ and $C\,\varepsilon\,\varepsilon''=\frac{\mathrm{d}}{\mathrm{d}x}(C\,\varepsilon\,\varepsilon')-C\,\varepsilon'^2$ can be integrated by parts. Upon neglecting the boundary terms, we obtain the approximation

$$\Phi^\ddagger[\varepsilon]=\int_{-\infty}^{+\infty}\frac{1}{2}\left(K(\chi)\,\varepsilon^2+\eta^2B^\ddagger(\chi)\,\varepsilon'^2\right)\mathrm{d}x\tag{6.9}$$

where

$$B^\ddagger(\chi)=B(\chi)-2\,C(\chi).\tag{6.10}$$

The quantity $B^\ddagger(\chi)$ is not necessarily positive, unlike the original gradient modulus $B(\chi)$ which is a diagonal entry of $\boldsymbol{t}(\chi)\geqslant 0$, see also (6.7). For $\chi=1$, for instance, $B^\ddagger(1)=-1.396\times10^{-2}k<0$, making the energy $\Phi^\ddagger[\varepsilon]$ in (6.9) non-positive. The set of positive values of $\chi$ that make $B^\ddagger(\chi)$ positive can be worked out as $\chi\in(0,0.602)$: we will explain in Section 6.4 why $B^\ddagger(\chi)$ has to be positive in the limit $\chi\to 0$ where the lattice becomes floppy.

## 6.3 Cholesky-based truncation preserving positivity

We restart from the homogenized energy given in (6.4), and apply Cholesky decomposition from Section 5 to the matrix $\boldsymbol{t}$ appearing in (6.5). Since there is only one scalar macroscopic strain measure $\varepsilon$ in the one-dimensional case, we can ignore the block structure of $\boldsymbol{t}$ which we assumed in Section 5 for the sake of generality.

Setting $\boldsymbol{\varepsilon}^{(0)}=(\varepsilon)$, $\boldsymbol{\varepsilon}^{(1)}=(\varepsilon')$, $\boldsymbol{\varepsilon}^{(2)}=(\varepsilon'')$, Equations (5.4–5.5) yield an alternative expression of the energy which is accurate to order $\eta^2$ included,

$$\Phi_+[\varepsilon]=\int_{-\infty}^{+\infty}\left(\frac{1}{2}D_0(\chi)\left\{\varepsilon(x)+\eta L_1(\chi)\,\varepsilon'(x)+\eta^2L_2(\chi)\,\varepsilon''(x)\right\}^2+\frac{\eta^2}{2}D_1(\chi)\,\varepsilon'^2(x)\right)\mathrm{d}x.\tag{6.11}$$

The homogenized moduli $D_i(\chi)$ and coefficients $L_i(\chi)$ are read off from Table 5.1 as

$$\begin{aligned}D_0(\chi)&=K(\chi) & L_1(\chi)&=\frac{A(\chi)}{K(\chi)}\\ D_1(\chi)&=B(\chi)-\frac{A^2(\chi)}{K(\chi)} & L_2(\chi)&=\frac{C(\chi)}{K(\chi)}.\end{aligned}\tag{6.12}$$

Combining (6.11) and (6.12), we obtain the positive, order-$\eta^2$-accurate expression of the energy for the pantograph as

$$\Phi_+[\varepsilon]=\int_{-\infty}^{+\infty}\left(\frac{1}{2}K(\chi)\left\{\varepsilon(x)+\eta\frac{A(\chi)}{K(\chi)}\varepsilon'(x)+\eta^2\frac{C(\chi)}{K(\chi)}\varepsilon''(x)\right\}^2+\frac{\eta^2}{2}\left(B(\chi)-\frac{A^2(\chi)}{K(\chi)}\right)\varepsilon'^2(x)\right)\mathrm{d}x.\tag{6.13}$$

In the non-symmetric case, the generalized strain $\{\ldots\}$ includes a correction of order $\eta$ proportional to $\varepsilon'$.

In the finite-contrast case, $\chi > 0$, the positivity of the energy $\Phi_+$ follows from (6.7), see also the general proof in Section 5.2. The limit of large elastic contrast, $\chi \to 0$, is studied next.

## 6.4 Floppy limit, $\chi \to 0$

In this section, we discuss the form of the energy $\Phi_+$ in (6.13) in the limit of large elastic contrast, $\chi \to 0$, where a floppy mode coupled to macroscopic extension appears. Specifically, we explain why the naïve procedure from Section 6.2 produces a positive energy in this particular case, $B^\ddagger(0) \geqslant 0$.

From the expressions of $K(\chi)$, $A(\chi)$ and $C(\chi)$ appearing in the second row of Table 6.1, on can obtain the following equivalents in the limit $\chi \to 0$

$$K(\chi) \sim 4k\,\chi, \qquad A(\chi) \sim \frac{2k}{69}\chi, \qquad C(\chi) = \frac{383\,k}{28\,566}\chi, \tag{6.14}$$

whereas $B(\chi)$ converges to a positive constant

$$B(\chi) \to \frac{8k}{69}. \tag{6.15}$$

The scaling behavior observed in (6.14–6.15) can be explained as follows. For $\chi = 0$, the diagonal matrix of microscopic moduli $\boldsymbol{k}(0)$ in (6.3) has a zero entry in the last, 13th, position on the diagonal and the first 12 components of the vector $\boldsymbol{F}_0(0)$ vanish, implying $\boldsymbol{k}(0)\cdot\boldsymbol{F}_0(0) = 0$. Now, from (6.6), the homogenized constants $K(\chi)$, $A(\chi)$ and $C(\chi)$ appear to be the dot product of $\boldsymbol{k}(\chi)\cdot\boldsymbol{F}_0(\chi) \to 0$ by one of the $\boldsymbol{F}_i(\chi)$'s, but $B(\chi)$ is not of this form. This is what makes $K(\chi)$, $A(\chi)$ and $C(\chi)$ proportional to $\chi$ in the limit $\chi \to 0$, whereas $B(\chi)$ has a non-zero limit. This behavior appears to be generic in structures possessing a soft mode, which is such that $\boldsymbol{k}\cdot\boldsymbol{F}_0$ is not full-rank, see [DLSS22].

In the limit $\chi \to 0$, the ratios $A/K$ and $C/K$ in (6.13) are therefore finite, whereas both $K$ and $A^2/K$ go to zero. In view of (6.13), the homogenized energy relevant to the soft limit is therefore

$$\Phi_+[\varepsilon] = \int_{-\infty}^{+\infty} \frac{\eta^2}{2} B(0)\,\varepsilon'^2(x)\,\mathrm{d}x. \tag{6.16}$$

On the other hand, the energy produced by the naïve truncation method in (6.9) becomes, in the limit,

$$\Phi^\ddagger[\varepsilon] = \int_{-\infty}^{+\infty} \frac{\eta^2}{2} B^\ddagger(0)\,\varepsilon'^2\,\mathrm{d}x. \tag{6.17}$$

Equation (6.10) shows that $B^\ddagger(0) = B(0)$ as $C(0) = 0$ by $(6.14)_3$. The energies (6.16) and (6.17) produced by the naïve and Cholesky methods therefore coincide in the presence of a soft mode, when the regular, leading-order elasticity is degenerate. This explains why the naïve truncation method predicts a correct (positive) strain-gradient elasticity in this soft limit.

# 7 Extension to dimension reduction: bending of an elastic slab

In this section we show that our approach extends naturally to dimension reduction. We analyze the bending of a linearly elastic slab in dimension 2. Specifically, we revisit the work of [LA20] and derive a second-order energy for the bending of an infinitely long rectangular 2D beam, fixing the non-positivity of the gradient-of-curvature modulus reported in this previous work.

## 7.1 Summary of dimension reduction

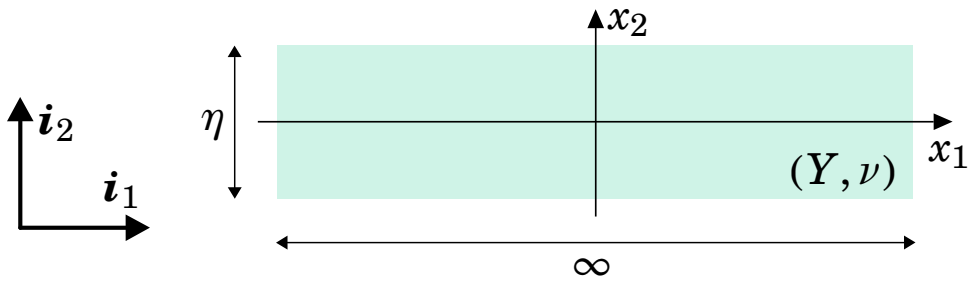

**Figure 7.1.** A linearly elastic slab in dimension 2 having Young modulus $Y$ and Poisson's ratio $\nu$. A second-order equivalent beam bending model is derived by dimension reduction, see Equation (7.13).

The rectangular domain of the slab is shown in Figure 7.1. It is infinite in the longitudinal $x_1$ direction and its width is denoted as $\eta$. We introduce the stretched transverse coordinate

$$y = \frac{x_2}{\eta}. \tag{7.1}$$

The stretched coordinate $y$ varies in the range $-\frac{1}{2} \leqslant y \leqslant +\frac{1}{2}$.

Dimension reduction addresses the limit $\eta \ll 1$, see [LA20], and can be summarized as follows. The displacement field relevant to bending is of the form (see Section 4.2 of [LA20])

$$\boldsymbol{u}(x_1, y) = [-\eta\, y\, V'(x_1) + \eta^2\, \xi_1(x_1, y)]\,\boldsymbol{i}_1 + [V(x_1) + \eta^2\, \xi_2(x_1, y)]\,\boldsymbol{i}_2, \tag{7.2}$$

where $V(x_1)$ denotes the macroscopic displacement in the transverse direction, $\xi_i(x_1,y)$ are the components of the microscopic displacement and $(\boldsymbol{i}_1,\boldsymbol{i}_2)$ is the Cartesian basis. The latter is subject to the constraint

$$\forall x_1, \quad \int_{-1/2}^{+1/2} \xi_i(x_1,y)\,\mathrm{d}y = 0 \tag{7.3}$$

warranting that the macroscopic displacement $V(x_1)\,\boldsymbol{i}_2$ is the average of the displacement $\boldsymbol{u}(x_1,y)$ in any particular cross-section with coordinate $x_1$.

Using asymptotic analysis, the microscopic displacement that minimizes the elastic energy for a prescribed deflection profile $V$ has been obtained in [LA20] in the form

$$\begin{array}{lcl} \xi_1(x_1,y) &=& 0 \quad - \; y\left(\dfrac{6+5\nu}{24}-\dfrac{2+\nu}{6}y^2\right)\eta\,\kappa'(x_1) \; + \cdots \\ \xi_2(x_1,y) &=& \nu\,\kappa(x_1)\left(\dfrac{y^2}{2}-\dfrac{1}{24}\right) \; + \quad 0\times\eta \quad + \cdots \end{array} \tag{7.4}$$

where $\kappa(x_1)$ denotes the macroscopic bending strain,

$$\kappa(x_1) = V''(x_1), \tag{7.5}$$

and $\nu$ is the Poisson's ratio of the material appearing in Equation (7.8) below. The solution (7.4) satisfies the constraint (7.3).

The microscopic 2D strain $(\nabla_{\boldsymbol{x}}\boldsymbol{u}+\nabla_{\boldsymbol{x}}\boldsymbol{u}^T)/2$ is obtained from (7.2) in the form $\eta\,\boldsymbol{E}(x_1,y)$, where

$$\boldsymbol{E}(x_1,y) = \left(-y\,\kappa(x_1)+\eta\frac{\partial\xi_1}{\partial x_1}\right)\boldsymbol{i}_1\otimes\boldsymbol{i}_1+\left(\frac{\partial\xi_1}{\partial y}+\eta\frac{\partial\xi_2}{\partial x_1}\right)\frac{\boldsymbol{i}_1\otimes\boldsymbol{i}_2+\boldsymbol{i}_2\otimes\boldsymbol{i}_1}{2}+\frac{\partial\xi_2}{\partial y}\boldsymbol{i}_2\otimes\boldsymbol{i}_2. \tag{7.6}$$

Inserting (7.4), we obtain

$$\begin{aligned} \boldsymbol{E}(x_1,y) = &\left[-y\,(\boldsymbol{i}_1\otimes\boldsymbol{i}_1-\nu\,\boldsymbol{i}_2\otimes\boldsymbol{i}_2)\right]\kappa(x_1) \\ &+\left[-(1+\nu)\left(\frac{1}{4}-y^2\right)\frac{\boldsymbol{i}_1\otimes\boldsymbol{i}_2+\boldsymbol{i}_2\otimes\boldsymbol{i}_1}{2}\right]\eta\,\kappa'(x_1) \\ &+\left[-y\left(\frac{6+5\nu}{24}-\frac{2+\nu}{6}y^2\right)\boldsymbol{i}_1\otimes\boldsymbol{i}_1+\cdots\right]\eta^2\,\kappa''(x_1) \\ &+\ldots \end{aligned} \tag{7.7}$$

The quantities appearing in square brackets in (7.7) have been denoted as $\boldsymbol{F}_i$ earlier, see for instance Equation (2.16). In the fourth line of (7.7), the ellipsis stands for a contribution coming from the order-$\eta^2$ term in the displacement, that has not been resolved in (7.4). This contribution does *not* contribute to the energy at order-$\eta^2$, however, by the argument already given in Remark 2.1.

The linear theory of elasticity in 2D yields the strain energy of the slab in the form

$$\Phi = \frac{\eta^3 Y}{2}\int_{-\infty}^{+\infty}\int_{-1/2}^{+1/2}\left(\frac{\nu}{1-\nu^2}\mathrm{tr}^2\boldsymbol{E}(x_1,y)+\frac{1}{1+\nu}\boldsymbol{E}(x_1,y):\boldsymbol{E}(x_1,y)\right)\mathrm{d}x_1\,\mathrm{d}y, \tag{7.8}$$

where $Y$ and $\nu$ are the Young modulus and Poisson's ratio, respectively, in the sense of two-dimensional elasticity. By using the identity $\boldsymbol{E}:\boldsymbol{E}=\boldsymbol{E}^D:\boldsymbol{E}^D+\frac{1}{2}\mathrm{tr}^2\boldsymbol{E}$, where $\boldsymbol{E}^D=\boldsymbol{E}-\frac{\boldsymbol{I}_2}{2}\mathrm{tr}\,\boldsymbol{E}$ is the deviatoric part, one can show that the range of elastic parameters corresponding to a stable material are

$$Y>0 \quad\text{and}\quad -1<\nu<1. \tag{7.9}$$

Note that the incompressible limit corresponds to $\nu=1$ in 2D, and not to $\nu_{3\mathrm{D}}=1/2$ as in 3D.

The prefactor $\eta^3$ in (7.8) comes from the fact that the actual 2D strain is $\eta\,\boldsymbol{E}$ (and not just $\boldsymbol{E}$) and that we are using the stretched coordinate $y$ in the transverse integration. For the energy to be finite in the limit $\eta\to 0$, we can conveniently assume that the Young modulus scales as $Y=\eta^{-3}\overline{Y}$, where $\overline{Y}=\mathcal{O}(1)$.

Inserting the expression (7.7) of the strain into (7.8) and integrating with respect to the transverse coordinate $y$, we obtain

$$\Phi^\star[\kappa] = \frac{\overline{Y}/12}{2}\int_{-\infty}^{+\infty}\left(\kappa^2(x_1)+\frac{2}{5}\left(1+\frac{11\,\nu}{12}\right)\eta^2\,\kappa(x_1)\,\kappa''(x_1)+\frac{1+\nu}{5}\eta^2\,\kappa'^2(x_1)\right)\mathrm{d}x_1 \tag{7.10}$$

where $\overline{Y}/12$ is the scaled bending modulus predicted by classical dimension reduction. Due to the left-right symmetry, there is no energy contribution coupling $\varepsilon$ and $\varepsilon'$ ($A$ terms).

The form $\Phi^\star=\frac{1}{2}\int\left(K\,\kappa^2+\eta^2\left(2\,C\,\kappa\,\kappa''+B\,\kappa'^2\right)\right)\mathrm{d}x_1$ of the energy matches that relevant to homogenization, see for instance (6.8) for the pantograph. The coefficients can be identified from (7.10) as

$$\begin{aligned} K &= \frac{\overline{Y}}{12} \\ C &= \frac{K}{5}\left(1+\frac{11\,\nu}{12}\right) \\ B &= \frac{K}{5}(1+\nu). \end{aligned} \tag{7.11}$$

By the same arguments as earlier, the homogenized moduli $K>0$ and $B\geqslant 0$ are positive over the range (7.9) of permissible elastic parameters. This is a consequence of the positivity of the energy which we started from in (7.8). The positivity of $C$ is coincidental.

### 7.2 Naïve truncation of the energy

Upon integration by parts of the $\kappa\,\kappa''$ term in (7.10) and after discarding the boundary terms, we recover the non-positive functional

$$\Phi^{\ddagger}[\kappa] = \frac{\overline{Y}/12}{2}\int_{-\infty}^{+\infty}\left(\kappa^2(x_1) - \frac{1}{5}\left(1+\frac{5\,\nu}{6}\right)\kappa'^2(x_1)\right)\mathrm{d}x_1, \tag{7.12}$$

which was the final result of [LA20], see their Equation [4.10]. The expression $B^{\ddagger} = B - 2\,C = -\frac{K}{5}\left(1+\frac{5}{6}\nu\right)$ of the *apparent* gradient modulus is similar to that derived earlier in the context of homogenization, see (2.29) and (6.10). The apparent modulus is negative, $B^{\ddagger} < 0$, which makes $\Phi^{\ddagger}$ non-convex.

### 7.3 Truncation preserving positivity

Applying instead the method from Section 2.6 to the one-dimensional energy $\Phi^{\star}$ in (7.10), we get

$$\Phi_{+}[\kappa] = \frac{\overline{Y}/12}{2}\int_{-\infty}^{+\infty}\left(\left\{\kappa(x_1) + \frac{1}{5}\left(1+\frac{11\,\nu}{12}\right)\eta^2\kappa''(x_1)\right\}^2 + \frac{1+\nu}{5}\eta^2\kappa'^2(x_1)\right)\mathrm{d}x_1. \tag{7.13}$$

This is a convex, second-order-accurate approximation of $\Phi^{\star}$. It is akin to the other energy functionals $\Phi_{+}$ derived earlier, see for instance in (2.34) for the laminate.

**Remark 7.1. (geometric compatibility)** The generalized strain appearing in (7.13) can be rewritten as $\left\{\kappa(x_1) + \frac{1}{5}(\cdots)\,\eta^2\kappa''(x_1)\right\} = \hat{V}''(x_1)$ in terms of the *alternate* macroscopic displacement $\hat{V}$,

$$\hat{V}(x_1) = V(x_1) + \frac{1}{5}\left(1+\frac{11\,\nu}{12}\right)\eta^2 V''(x_1). \tag{7.14}$$

In this sense, the generalized bending strain is geometrically compatible, as is any strain measure in dimension 1. This yields a compact and positive expression of the homogenized energy in the form

$$\frac{\overline{Y}/12}{2}\int_{-\infty}^{+\infty}\left(\hat{V}''^2(x_1) + \frac{1+\nu}{5}\eta^2\hat{V}'''^2(x_1)\,(x_1)\right)\mathrm{d}x_1, \tag{7.15}$$

which differs from the energy $\Phi_{+}[\kappa]$ in (7.13) by negligible terms, of order $\eta^4$, coming from subdominant contributions to the strain gradient.

## 8 Discussion and conclusion

We propose a truncation method for energy functionals produced by higher-order homogenization. It solves a long-standing issue concerning the presence of non-positive strain-gradient stiffnesses. This issue made simulations of homogenized models ill-posed and may explain the limited popularity of high-order homogenization. In our approach, the non-positive, higher-order homogenized constants are revisited as *coefficients* entering in the definition of generalized strain measures that are used as arguments of a *positive* quadratic form. The positivity of the energy $\Phi_{+}$ obtained in this way follows from the positivity of the initial microscopic energy. The truncation preserves the order of accuracy of the homogenization procedure.

The positive functional $\Phi_{+}$ produced by truncation depends solely and uniquely on the homogenized constants that are obtained by homogenization at a given order $p$. The truncation method can be implemented as post-processing step that takes as an input the order-$p$ Taylor expansion $\Phi^{\dagger}$ of the energy produced by conventional homogenization, such as that in Equations (2.26) and (6.8). To derive the positive energy $\Phi_{+}$ for the laminate which we obtained in (2.34), for example, we could simply extract the non-positive energy functional $\Phi^{\dagger}$ in (2.26) from previous work [Le15] (carefully avoiding the alternate form (2.28) that involves an additional integration by parts) and apply the decomposition method directly.

Thanks to the iterative nature of the underlying Cholesky decomposition, the positive functional $\Phi_{+}$ obtained at order $p$ refines that obtained at a lower order $p' < p$, by adding higher-order terms without modifying the existing terms. By extending the Cholesky factorization done in Section 5 at order $p' = 2$ to order $p = 4$, for instance, one obtains an energy of the form

$$\Phi = \frac{1}{2}\int\left(\hat{\boldsymbol{\varepsilon}}^{(0)}\cdot\boldsymbol{D}_0\cdot\hat{\boldsymbol{\varepsilon}}^{(0)} + \eta^2\,\hat{\boldsymbol{\varepsilon}}^{(1)}\cdot\underline{\underline{\boldsymbol{D}}}_1\cdot\hat{\boldsymbol{\varepsilon}}^{(1)} + \eta^4\,\hat{\boldsymbol{\varepsilon}}^{(2)}\cdot\underline{\underline{\underline{\underline{\boldsymbol{D}}}}}_2\cdot\hat{\boldsymbol{\varepsilon}}^{(2)} + \cdots\right)\mathrm{d}\boldsymbol{x} \tag{8.1}$$

where the generalized strains $\hat{\boldsymbol{\varepsilon}}^{(k)}$ are given by

$$\begin{array}{lcl}
\hat{\boldsymbol{\varepsilon}}^{(0)} &=& \boldsymbol{\varepsilon}^{(0)} + \eta\,\underline{\boldsymbol{L}}_1^T\cdot\boldsymbol{\varepsilon}^{(1)} + \eta^2\underline{\underline{\boldsymbol{L}}}_2^T\cdot\boldsymbol{\varepsilon}^{(2)} + \eta^3\underline{\underline{\underline{\boldsymbol{L}}}}_3^T\cdot\boldsymbol{\varepsilon}^{(3)} + \eta^4\underline{\underline{\underline{\underline{\boldsymbol{L}}}}}_4^T\cdot\boldsymbol{\varepsilon}^{(4)} + \ldots\\
\hat{\boldsymbol{\varepsilon}}^{(1)} &=& \boldsymbol{\varepsilon}^{(1)} + \eta\,\underline{\underline{\tilde{\boldsymbol{L}}}}_2^T\cdot\boldsymbol{\varepsilon}^{(2)} + \eta^2\underline{\underline{\underline{\tilde{\boldsymbol{L}}}}}_3^T\cdot\boldsymbol{\varepsilon}^{(3)} + \ldots\\
\hat{\boldsymbol{\varepsilon}}^{(2)} &=& \boldsymbol{\varepsilon}^{(2)} + \ldots
\end{array} \tag{8.2}$$

and $\boldsymbol{\varepsilon}^{(k)}$ represents the components of the tensor $\nabla^k\boldsymbol{\varepsilon}$ flattened into a vector. The homogenized constants having 2 underlines or less are identical to those obtained at order $p' = 2$, see Equations (5.4–5.5). Increasing the homogenization order to $p = 4$ brings in the quantities bearing 3 or 4 underlines, namely a new, positive block of homogenized moduli $\underline{\underline{\underline{\underline{\boldsymbol{D}}}}}_2$ acting on the second gradient of strain $\hat{\boldsymbol{\varepsilon}}^{(2)}$, and higher-order coefficients $\underline{\underline{\underline{\boldsymbol{L}}}}_3$, $\underline{\underline{\underline{\tilde{\boldsymbol{L}}}}}_2$, $\underline{\underline{\underline{\underline{\boldsymbol{L}}}}}_4$ and $\underline{\underline{\underline{\tilde{\boldsymbol{L}}}}}_3$ refining the generalized strain measures $\hat{\boldsymbol{\varepsilon}}^{(0)}$ and $\hat{\boldsymbol{\varepsilon}}^{(1)}$ that were already present at order 2.

With $\underline{\underline{L}}_1 \neq \underline{\underline{\tilde{L}}}_2$ and $\underline{\underline{L}}_2 \neq \underline{\underline{\tilde{L}}}_3$ in general, the second generalized strain $\hat{\boldsymbol{\varepsilon}}^{(1)}$ appearing in (8.2) is not simply the gradient of the first generalized strain $\hat{\boldsymbol{\varepsilon}}^{(0)}$.

Overall, the Cholesky-based truncation handles non-centro-symmetric structures, see Section 6, continuous as well as discrete structures, arbitrary space dimension, and arbitrary homogenization orders $p$, including *odd* ones—the case $p=3$, for instance, corresponds to discarding the quantities $\underline{\underline{D}}_2$, $\underline{\underline{L}}_4$ and $\underline{\underline{\tilde{L}}}_3$ in (8.1–8.2).

The truncation method restores positiveness by introducing a higher-order non-asymptotic term, which in the case of the laminate and at homogenization order $p=2$ takes the form $\frac{\eta^4}{2}\nabla^2\boldsymbol{\varepsilon}\cdot[\boldsymbol{C}^T\cdot\boldsymbol{K}^{-1}\cdot\boldsymbol{C}]\cdot\nabla^2\boldsymbol{\varepsilon}$, see (2.34). As a result, the order of the boundary-value problem for the displacement is 6, which is higher than the order (4) of the boundary-value problem produced by the naïve truncation method. In general, when homogenizing at an even order $p$, the boundary-value problem associated with the positive energy $\Phi_+$ is of order $2(p+1)$, which is 2 more than that produced by the naïve truncation method, namely $2p$. With this relatively high differentiation order, the numerical solution of the boundary-value problem produced by the positive functional $\Phi_+$ may require special attention.

In hindsight, at least two instances of positive energy functionals of the type $\Phi_+$ have appeared in earlier work on second-order dimension reduction:

- The one-dimensional energy governing the twisting of linearly elastic beams has been obtained in Equation [5.17] of [AL21] in a form which we can rewrite as

$$\Phi_+ = \int_0^L \frac{1}{2}\mu J\left(\{\tau(x)+\eta^2 L\,\tau''(x)\}^2+\eta^2 b\,\tau'^2(x)\right)\mathrm{d}x, \tag{8.3}$$

  where $\mu J>0$ is the classical twisting modulus, $\eta$ is the cross-section radius, $\{\tau+\ldots\}$ is a generalized twisting strain that corrects the classical twisting strain $\tau(x)=\theta'(x)$ by a higher-order term, $\theta(x)$ is the rotation of the cross-section about the axis, $L$ is a constant depending on the cross-section properties, and $b$ is a higher-order modulus. The reasoning done in Section 5.2 of the present work proves that the higher-order modulus $b>0$ is positive, as long as the underlying elastic material is stable. The positivity of $b$ had only been noted in the special case of elliptical cross-section in the original paper [AL21].

- The one-dimensional energy governing the stretching of an elastic cylinder subjected to *finite* deformations has been initially derived in [AH16] and later reformulated in Appendix B.1 of [AH19] in the form

$$\Phi_+ = \int_0^L W_0\left(\left\{\lambda(X)+\eta^2 L(\lambda(X))\,\lambda'^2(X)\right\}\right)\mathrm{d}X \tag{8.4}$$

  where $\lambda(X)=\mathrm{d}x/\mathrm{d}X$ is the effective longitudinal stretch, $X$ and $x$ are the longitudinal coordinates in reference and deformed configurations, respectively, $\eta$ is the initial radius of the cylinder, $W_0(\lambda)$ is the energy per unit length when the cylinder is subjected to uniform simple traction, the coefficient $L$ was found as $L(\lambda)=1/(16\lambda^4)$. This energy is similar to that obtained in (3.20) for the triangular truss, when the quadratic potential $\frac{1}{2}\boldsymbol{\varepsilon}:\boldsymbol{K}:\boldsymbol{\varepsilon}$ applicable to linear elasticity is identified with the non-quadratic potential $W_0(\lambda)$ relevant to finite-strain elasticity. By contrast with the linear case, however, the generalized strain $\hat{\lambda}=\{\lambda+\eta^2\lambda'^2/(16\lambda^4)\}$ used as an argument of $W_0$ in (8.4) includes a correction proportional to $(\eta\lambda')^2$ and not $\eta^2\lambda''$. The functional (8.4) will typically yield stable solution whenever the underlying one-dimensional potential $W_0(\lambda)$ is convex.

The energy functional (8.4) relevant to finite elasticity suggests that the proposed reduction method can be extended to *nonlinear* problems in homogenization and dimension reduction.

In the nonlinear analysis of the stretched cylinder [AL21], it has been noted that the generalized one-dimensional stretch $\hat{\lambda}(X)=\{\lambda(X)+\eta^2\lambda'^2(X)/(16\lambda^4(X))\}$ has a simple physical interpretation, unlike $\lambda(X)$: it is the average over the cross-section of the *microscopic* nonlinear measure of longitudinal strain. This suggests that the generalized strain measures produced by the Cholesky procedure, such as $\{\boldsymbol{\varepsilon}+\eta\cdots+\eta^2\cdots\}$, are more meaningful than the standard strain measures, such as $\boldsymbol{\varepsilon}=(\nabla\boldsymbol{u}+\nabla\boldsymbol{u}^T)/2$, used in previous work on higher-order homogenization. We hope that additional useful mathematical properties of the homogenized energy concerning, e.g., the regularity of the solution, can be identified by working with this alternate strain measure.

Having restored the crucial property of positivity, our method could help increasing the popularity of higher-order homogenization. There remains a last obstacle towards this goal, which will hopefully be clarified in future work: we have limited attention to infinite systems in this paper and it is yet unclear how one can account for boundary conditions in the energy approach.

*Acknowledgements*: we would like to thank Hussein Nassar for pointing out the connection of our work with the Cholesky method, and Antoine Gloria for insightful comments.

# Appendix A Notation for multiple contractions

Given two tensors $\boldsymbol{a}$ and $\boldsymbol{b}$ of respective order $p$ and $p'$, the contraction operations are defined as

$$\begin{aligned}
(\boldsymbol{a}:\boldsymbol{b})_{i_1\ldots i_{p-2}i'_3\ldots i'_p} &= a_{i_1\ldots i_{p-2}kl}\, b_{kli'_3\ldots i'_p} \quad ,\\
(\boldsymbol{a}\therefore\boldsymbol{b})_{i_1\ldots i_{p-3}i'_4\ldots i'_p} &= a_{i_1\ldots i_{p-3}klm}\, b_{klmi'_4\ldots i'_p} \quad ,\\
(\boldsymbol{a}::\boldsymbol{b})_{i_1\ldots i_{p-4}i'_5\ldots i'_p} &= a_{i_1\ldots i_{p-4}klmn}\, b_{klmni'_5\ldots i'_p} \quad .
\end{aligned} \tag{A.1}$$